\documentclass{article}
\usepackage[english]{babel}
\usepackage[utf8]{inputenc}
\usepackage{johd}
\usepackage{amsmath}
\usepackage{array}
\usepackage{graphicx}
\usepackage{caption}
\usepackage{booktabs}
\usepackage{amssymb}
\usepackage{textcomp}
\usepackage{subfig}
\usepackage{natbib}
\usepackage{rotating}
\usepackage[export]{adjustbox}
\usepackage{wrapfig}
\usepackage{sidecap}
\usepackage{siunitx}
\DeclareUnicodeCharacter{2212}{-}
\title{Multivariate Affine GARCH with Heavy Tails: A Unified Framework for Portfolio Optimization and Option Valuation}

\author{Ayush Jha$^{1}$$^{*}$, Abootaleb Shirvani$^{2}$, Ali Jaffri$^{1}$, Svetlozar T. Rachev$^{3}$ and Frank J. Fabozzi$^{4}$ \\
        \small $^{1}$Department of Economics, Texas Tech University \\
        \small $^{2}$Department of Mathematical Sciences, Kean University \\
        \small $^{3}$Department of Mathematics and Statistics, Texas Tech University \\
        \small $^{4}$Carey Business School, Johns Hopkins University \\\\
        \small $^{*}$Corresponding author: Ayush Jha; \tt{ayush.jha@ttu.edu} \\
}
\date{}

\begin{document}
\maketitle
\begin{abstract} 
\noindent This paper develops and estimates a multivariate affine GARCH(1,1) model with Normal Inverse Gaussian innovations that captures time-varying volatility, heavy tails, and dynamic correlation across asset returns. We generalize the Heston–Nandi framework to a multivariate setting and apply it to 30 Dow Jones Industrial Average stocks. The model jointly supports three core financial applications: dynamic portfolio optimization, wealth path simulation, and option pricing. Closed-form solutions are derived for a Constant Relative Risk Aversion (CRRA) investor’s intertemporal asset allocation, and we implement a forward-looking risk-adjusted performance comparison against Merton-style constant strategies. Using the model’s conditional volatilities, we also construct implied volatility surfaces for European options, capturing skew and smile features. Empirically, we document substantial wealth-equivalent utility losses from ignoring time-varying correlation and tail risk. These findings underscore the value of a unified econometric framework for analyzing joint asset dynamics and for managing portfolio and derivative exposures under non-Gaussian risks.
\end{abstract}

\noindent

\section{Introduction}
\setlength{\parskip}{10pt}

\noindent Understanding how volatility, tail risk, and cross-asset 
correlation evolve over time is critical to asset pricing, portfolio construction, and risk management. Yet, existing models often struggle to simultaneously capture the dynamic and non-Gaussian features of joint asset returns in a tractable, implementable framework. This paper addresses a central question in dynamic portfolio theory and financial econometrics: What is the economic cost of ignoring dynamic cross-asset volatility and heavy-tailed return behavior in multi-asset investment and derivatives pricing decisions?

\noindent A common hedging practice illustrates the problem. Investors frequently pair each risky position with an index fund or ETF to dampen systematic exposure. While intuitive, this bivariate approach fragments information: it measures each asset’s volatility in isolation and treats the hedge’s volatility as exogenous. When many such pairs reside in the same portfolio, shocks propagate through correlation channels that the pairwise filters cannot detect, distorting estimates of risk premia and generating sub-optimal rebalancing rules. When we simply aggregate up these one-asset-one-index strategies, the conditioning on each asset’s individual volatility–while ignoring its covariance with the rest of the portfolio–can lead to significant identification challenges in measuring true risk premia. 

\noindent Investors who manage large portfolios comprised of N risky holdings therefore need a more comprehensive approach that recognizes the dynamics of the joint distribution of asset returns. To answer this, we propose a multivariate affine GARCH(1,1) framework with Normal Inverse Gaussian (NIG) innovations. This generalizes the univariate \citet{hn2000} GARCH model to a high-dimensional setting, preserving closed-form variance dynamics and supporting both return forecasting and derivative valuation. By nesting affine structure within a heavy-tailed distributional setup, the model accommodates skewness, excess kurtosis, and evolving correlations—characteristics that empirical asset returns frequently display.

\noindent We estimate the model on a 30-stock panel drawn from the Dow Jones Industrial Average (DJIA) and the DJIA index itself, providing filtered estimates of variances and correlations in real time. Conditional volatility surfaces are then constructed using Heston–Nandi-based closed-form GARCH option pricing. In parallel, we implement a CRRA utility-based portfolio optimization to evaluate the wealth-equivalent losses of adopting simpler strategies—such as constant Sharpe ratio allocation—instead of dynamic risk-aware strategies implied by the model. The empirical analysis highlights the economic relevance of properly accounting for joint tail risk and volatility co-movement, particularly when estimating option prices and simulating long-term wealth paths.

\noindent Our study therefore delivers three main contributions. First, we introduce an empirically viable multivariate affine GARCH model with NIG innovations that unifies portfolio allocation, option valuation, and dynamic or forward-looking portfolio optimization. Second, we embed the filter in an intertemporal optimization problem and document sizeable WELs relative to the constant-variance benchmark. Third, we show that the resulting implied-volatility surfaces capture market-observed smiles and smirks that stem from correlation risk. Collectively, these results demonstrate that modeling joint, time-varying volatility is not a statistical nicety but a prerequisite for optimal portfolio and derivatives management.

\noindent The contributions of the paper are both theoretical and practical. Theoretically, we fill a structural gap in the literature by offering a multivariate GARCH specification that is affine, tractable, and analytically solvable for both optimal control and option valuation. Practically, we show that ignoring these features can lead to significant portfolio underperformance and option mispricing. The framework is flexible enough to support extensions to macro-finance, systemic risk monitoring, or joint modeling of equities and derivatives.

\noindent The remainder of the paper proceeds as follows. Section 2 reviews the relevant literature on affine GARCH models and dynamic portfolio choice. Section 3 presents the multivariate MHN–GARCH specification, describes the data, outlines the estimation strategy, and develops the option-pricing framework. Section 4 reports empirical results, including optimal allocations, efficient frontiers, implied-volatility surfaces, and WEL calculations. Section 5 concludes and offers directions for future research.

\section{Literature Review} \label{lit}

\noindent This paper contributes to the growing literature at the intersection of financial econometrics, dynamic asset allocation, and derivative pricing. Our model extends and unifies insights from several key strands of research on time-varying volatility, multivariate risk modeling, and dynamic control-based portfolio management.

\noindent A foundational contribution in this domain is \citet{hn2000}, who derive a closed-form solution for option valuation under a univariate GARCH(1,1) process with an affine variance structure. Our work generalizes this framework to a multivariate setting, allowing for time-varying variances, evolving correlations, and heavy-tailed innovations via the Normal Inverse Gaussian (NIG) distribution. This extension enables consistent modeling of multiple assets alongside a reference index such as the DJIA.

\noindent \citet{AEM2008} demonstrate the empirical viability of GARCH-based option pricing with asymmetric filters and non-Gaussian errors, while \citet{Stentoft} applies the NIG distribution in GARCH settings for American option valuation. Our multivariate specification builds upon these advances by offering joint estimation of return dynamics, conditional variances, and co-movement structures in a tractable framework that preserves the closed-form benefits of Heston–Nandi.

\noindent The literature has evolved from comparing GARCH and stochastic volatility models (\citet{Lss2002}; \citet{Chenlueng2005}), toward embedding richer distributional features such as skewness and kurtosis (\citet{MennRachev2005}; \citet{Chorro2012}). More recently, \citet{CZ2022} assess GARCH specifications by their ability to reproduce VIX behavior, highlighting the importance of model structure for capturing second-moment dynamics. Our approach complements these efforts by introducing a joint affine structure that accounts for cross-asset variance and correlation dynamics in both pricing and allocation contexts.

\noindent Incorporating correlation shocks and volatility spillovers into multivariate frameworks has received growing attention. \citet{KLY2014} incorporates VIX-based state vectors into GARCH filters, while \citet{Bergen2018} and \citet{escobar2025} introduce fully multivariate affine GARCH models for portfolio optimization under uncertainty. We expand upon these contributions by embedding the MHN–GARCH model in an intertemporal utility maximization problem and computing wealth-equivalent losses (WEL) from using simpler, static strategies.

\noindent Beyond option pricing, this paper also connects to dynamic portfolio theory under stochastic volatility. \citet{Merton1969} and \citet{Merton1971} laid the foundation for continuous-time intertemporal optimization under uncertainty. Subsequent discrete- and continuous-time models (\citet{BrennanSchwartzLagnado1997}; \citet{CampbellViceira1999}; \citet{Barberis2000}; \citet{Viceira2001}; \citet{Wachter2002}) highlight how mean-reverting returns and nontradable risks shape optimal asset allocation. \citet{BrennanXia2002} incorporate inflation as an additional state variable, and \citet{CampbellViceira2002} embed a vector‑autoregressive structure to capture multivariate predictability. International and regime‑switching dimensions are introduced by \citet{AngBekaert2002}, who show that latent volatility regimes materially affect global hedge demands. Our discrete-time CRRA optimization framework aligns with these traditions but updates them by incorporating filtered volatilities and co-dependencies from a multivariate affine GARCH system.

\noindent Taken together, this literature motivates our integrated modeling of return dynamics, forward-looking asset allocation, and derivative pricing. Our MHN–GARCH specification fills a structural gap: it offers analytical tractability and empirical feasibility for high-dimensional settings where volatility clustering, tail risk, and cross-asset correlation evolve jointly.

\section{Methodology: Modeling Joint Volatility and Correlation Dynamics} \label{model}

\noindent Accurately modeling the joint dynamics of volatility, correlation, and risk premia is essential for portfolio optimization and derivative pricing. Traditional GARCH models, though widely used, often fall short in two critical respects: they are typically univariate, ignoring inter-asset dependencies, and they assume Gaussian innovations, which fail to capture the heavy tails and skewness observed in asset returns. This section introduces a multivariate Heston–Nandi GARCH (MHN–GARCH) model with Normal Inverse Gaussian (NIG) innovations, designed to address both limitations in a unified framework.

\noindent The MHN–GARCH model generalizes the well-known \citet{hn2000} framework to a multivariate setting while preserving its analytical tractability. Each asset’s return is modeled jointly with a market index, allowing us to capture both idiosyncratic and systematic volatility dynamics. The use of NIG innovations enables the model to capture asymmetric, heavy-tailed behavior, enhancing its capacity to reflect empirical return distributions and to generate realistic option-implied volatility surfaces. Unlike dynamic conditional correlation (DCC) GARCH, Baba-Engle-Kraft-Kroner (BEKK) GARCH, or other high-dimensional GARCH variants that require either restrictive assumptions or intensive computation, the MHN–GARCH model features a filtered, closed-form updating rule for variances, covariances, and conditional means. This tractability allows for seamless integration into dynamic portfolio allocation problems and option pricing tasks, while avoiding overfitting or identification problems common to large multivariate systems.

\noindent To operationalize the model, we construct bivariate index–stock pairs across all 30 Dow Jones Industrial Average (DJIA) constituents and the DJIA index itself. This design reflects real-world hedging practices while capturing cross-asset interactions. We then estimate the model via maximum likelihood, filtering joint volatility paths and computing dynamic mean returns as a function of conditional variances and risk premia.
The resulting parameter estimates allow for period-by-period computation of variance–covariance matrices, expected returns, and co-market prices of risk. These filtered quantities feed directly into two applications developed later in the paper: closed-form option valuation and CRRA-based intertemporal asset allocation with Wealth Equivalent Loss (WEL) metrics.
This section proceeds by formally specifying the MHN–GARCH process, detailing its estimation via maximum likelihood, and outlining how the model enables forward-looking portfolio construction and option pricing under non-Gaussian dynamics.

\subsection{Data}

\noindent The empirical analysis draws on daily return data from the 30 constituents of the Dow Jones Industrial Average (DJIA), along with the DJIA index itself, resulting in a 31-asset universe. Price data were obtained from Bloomberg Terminal on November 20, 2024, corresponding to the official composition of the index on that date. The sample spans from November 23, 2020, through November 18, 2024, encompassing approximately 1,008 trading days. We use unadjusted closing prices, recorded prior to the ex-dividend date, to maintain consistency with standard option-pricing conventions. Consequently, our computed returns do not incorporate dividend effects, focusing instead on pure price dynamics.

\noindent To construct the joint volatility and correlation framework necessary for our multivariate model, each stock is paired with the index to form a bivariate return system. This design allows us to account for co-movements between individual assets and the broader market, aligning with real-world practices in risk management and hedging. By capturing idiosyncratic and systematic shocks in a unified bivariate setting, this structure provides a realistic basis for simulating joint risk exposures and designing dynamic portfolios.

\noindent We transform price series into continuously compounded log returns to ensure stationarity and to satisfy the statistical assumptions required for GARCH-type modeling. The resulting return series exhibit time-varying volatility, excess kurtosis, and dynamic correlation—features that motivate the use of a multivariate affine GARCH process with heavy-tailed innovations.
This dataset supports the empirical implementation of the MHN–GARCH model in two primary domains: (1) dynamic asset allocation for a CRRA investor, and (2) option valuation using Heston–Nandi closed-form expressions calibrated on filtered volatilities. The next section formalizes the MHN–GARCH model structure used to extract conditional variance–covariance matrices and price dynamics.

\subsection{MHN-GARCH (1,1) Process} \label{MHN-GARCH Process}

\noindent To capture the dynamics of time-varying volatility and correlation in asset returns, we introduce a multivariate affine GARCH(1,1) framework that builds upon the \citet{hn2000} model. Our specification generalizes this well-known univariate structure to a bivariate setting for each stock–index pair, while incorporating NIG innovations to better reflect empirical features such as skewness and excess kurtosis. This construction allows us to model the joint evolution of conditional variances and correlations with a high degree of flexibility, while preserving closed-form tractability for both portfolio optimization and option pricing applications. The model is designed to deliver filtered estimates of volatility and co-movement that are consistent with observed return behavior, and to support forward-looking risk assessments under non-Gaussian dynamics.

\noindent Let $(\Omega, \mathcal{F}, \mathbb{P})$ denote a probability space equipped with the discrete filtration $\{\mathcal{F}_{t}\}_{t=0}^{T}$.  We model the log prices of $N$ risky assets using the multivariate Heston–Nandi framework of \citet{hn2000}, extended analytically by \citet{escobar2025}. Let the log of the spot price of the risky asset $i$ at time $t$ be $log(S_{it}) = X_{it}$, which follows the following process in discrete time: 

\begin{align}
X_{n,t} &= X_{n,t-1} + r + \lambda_{1,n}\, h_{1,t} + \lambda_n\, h_{n,t} + a_{n}\, \sqrt{h_{1,t}}\, z_{1,t} + \sqrt{h_{n,t}}\, z_{n,t},\;\text{where}\;X_{n,0}(0) = x_{n,0} > 0,\;\;n=1,2,\dots, \label{log-prices}
\end{align}

\medskip

\noindent where $r$ is the risk--free rate and the conditional variances evolve according to
\medskip
\begin{align}
   h_{i,t} = \omega_i + \beta_i\, h_{i,t-1} + \alpha_i \left( z_{i,t-1} - \theta_i\, \sqrt{h_{i,t-1}} \right)^2,\quad i=1,\dots,n, \label{variance}
\end{align}

\medskip
\noindent $h_{i,t}$ is the conditional volatility of asset $i$'s log return between time $t$ and $t-1$ and $z_{i,t}$ are the standardized i.i.d innovations that follow an NIG process. \noindent Stationarity requires the familiar restriction $\beta_i + \alpha_i \theta_i^{2} < 1$, which implies an unconditional variance $\bar{h}_i = (\omega_i + \alpha_i)/(1 - \beta_i - \alpha_i \theta_i^{2})$.  The covariance between next‑period variance and the current log price for assets 1 and $n$ is  

\medskip

\begin{align*}
\operatorname{Cov}_{t-1}[h_{1,t+1},X_{1,t}] &= -2\alpha_1 \theta_1 h_{1,t}, \\[2pt]
\operatorname{Cov}_{t-1}[h_{n,t+1},X_{n,t}] &= -2\alpha_n \theta_n h_{n,t}\,(a_n+1),\;\;n=1,2,\dots,
\end{align*}

\medskip

\noindent highlighting how volatility news feeds back into returns in the presence of the cross‑asset coefficient $a_n$.

\noindent The variance equations specified below follow an autoregressive structure similar to standard GARCH(1,1) models, but are adapted to maintain an affine form. This ensures that the conditional variance evolves in a statistically robust and analytically convenient way, particularly for computing the log-likelihood and deriving option pricing solutions. The use of affine GARCH dynamics enables direct interpretation of how past shocks and variance levels influence future volatility, and allows for recursive computation of the variance path during estimation and simulation.

\begin{figure}[htbp]
    \centering
    \includegraphics[width=\textwidth,%
      height=\textheight,%
      keepaspectratio]{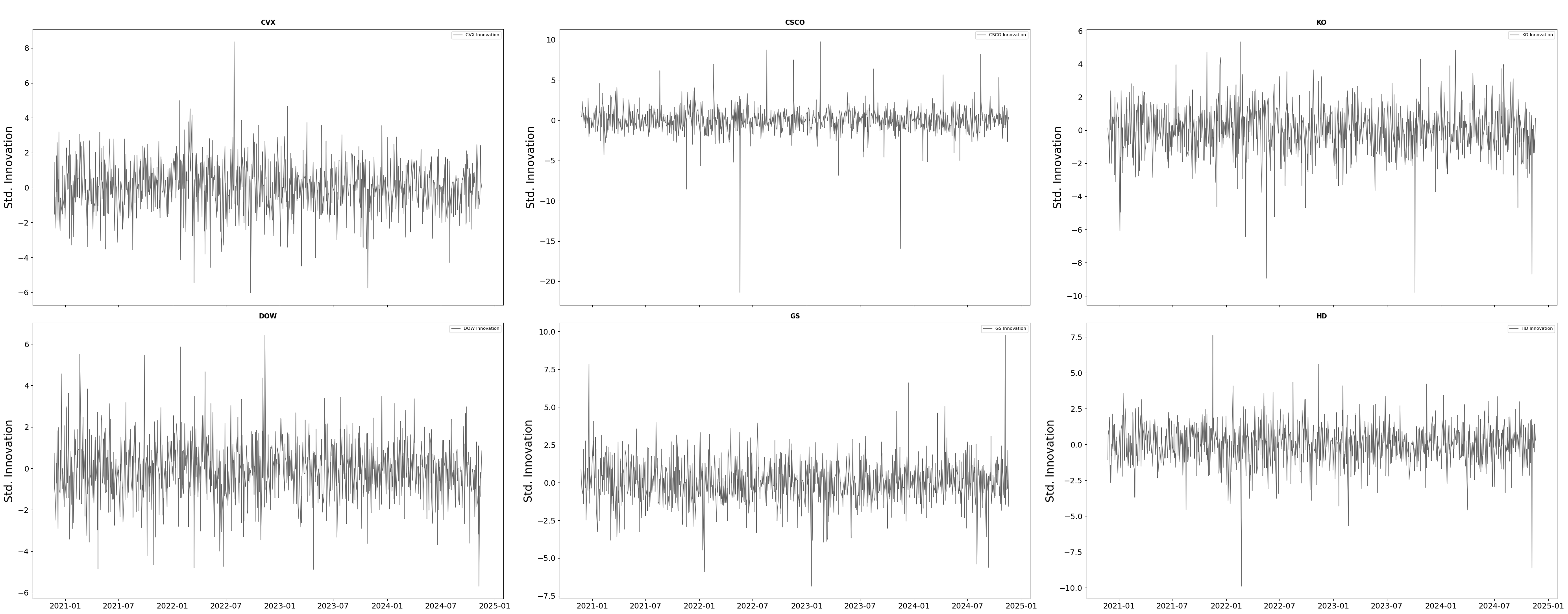}
    \caption{NIG innovations (asset‑specific unobservables)}
    \label{fig:innovations}
\end{figure}

\noindent Figure~\ref{fig:innovations} presents the standardized innovations obtained from the MHN‑GARCH(1,1) estimation\footnote{We plot the NIG innovations for a sample of six assets in the portfolio to demonstrate the nature of unobservables in this model. Other asset innovations are in appendix~\ref{innov}.}. Each sequence reflects the asset‑specific disturbance extracted after jointly filtering the asset with its benchmark index, thereby disentangling idiosyncratic shocks from the systematic component embedded in the index. The bivariate filtration captures nonlinear dependence and pronounced deviations from Gaussianity that would be obscured under marginal (univariate) specifications. Identification is strengthened by the correlation‑driver coefficient~$a_{n}$, which enters the log‑returns and innovation processes. This parameter preserves orthogonality between index‑level and asset‑level shocks while allowing cross‑asset volatility to propagate consistent with observed Dow‑30 dynamics, ensuring that the model represents the joint risk structure relevant for optimal portfolio allocation.

\begin{figure}[htbp]
    \centering
    \includegraphics[width=\textwidth,%
      height=\textheight,%
      keepaspectratio]{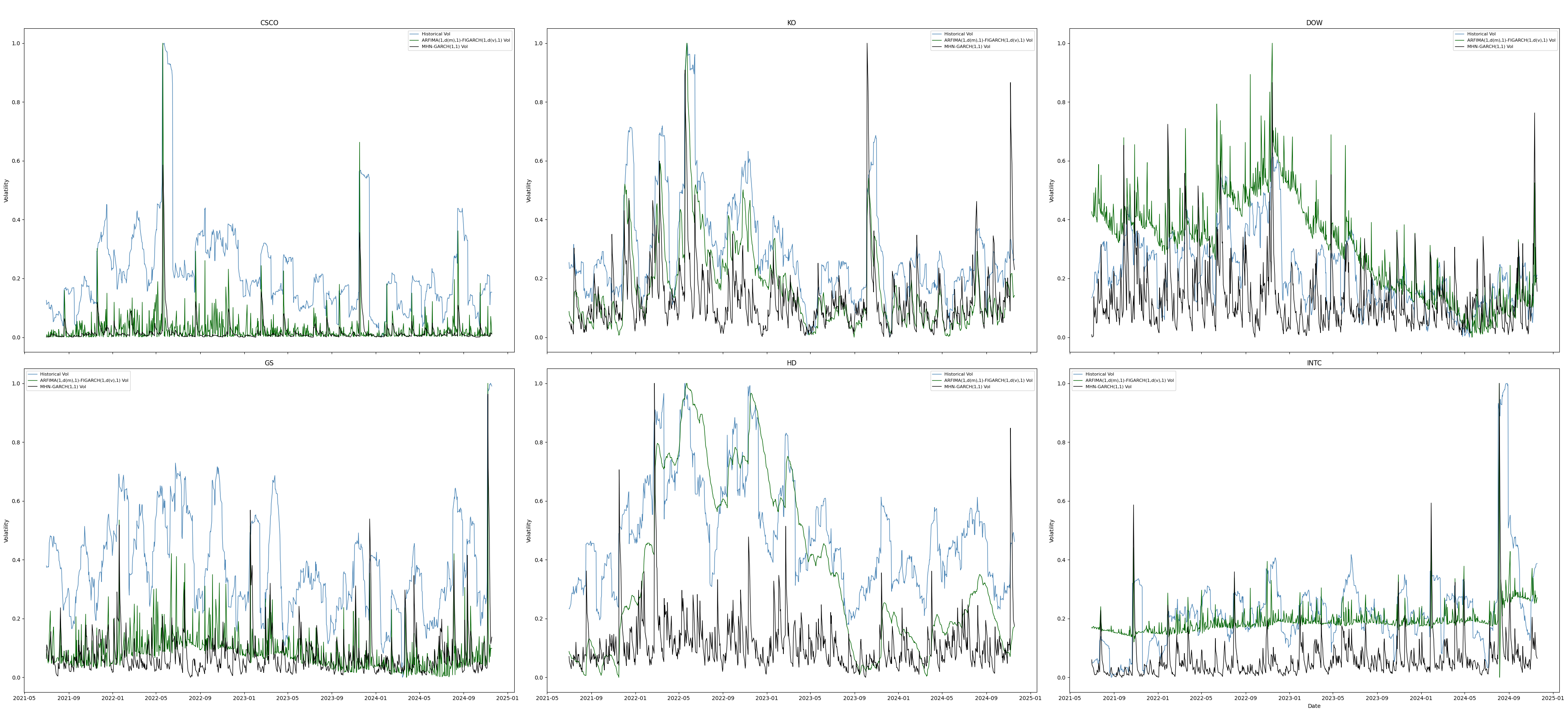}
    \caption{Model-based Volatility}
    \label{fig:vol_1}
\end{figure}

\noindent In figure~\ref{fig:vol_1}, we plot the evolution of three volatility estimates for a sample of 6 assets in the Dow Jones Industrial Average (DJIA): historical volatility, a univariate ARFIMA(1,$d(m)$,1)–FIGARCH(1,$d(v)$,1) conditional volatility, and the MHN-GARCH(1,1) conditional volatility\footnote{Appendix~\ref{vol} plots other constituents of the DJIA. All volatility series are normalized with mean 0 and variance 1 to match the scale of asset variance in all models.}. Each subplot focuses on a single stock’s time series. By comparing these three volatility estimates, we aim to illustrate how different modeling assumptions, historical versus univariate or multivariate parametric frameworks, can yield distinct volatility dynamics with respect to parameter sensitivity.

\noindent The historical volatility measure is computed by taking the rolling standard deviation of daily log returns over a fixed window (20 trading days). This empirical approach directly reflects recent historical price fluctuations, serving as a benchmark for short-run historical variability. However, it can be noisy: abrupt market shocks within a short window may cause pronounced spikes, while large outliers or a cluster of extreme observations can keep volatility elevated longer than the underlying volatility might suggest. Historical volatility does not impose a forward-looking structure or leverage cross-sectional information from other assets.

\noindent We next estimate a univariate ARFIMA(1,$d(m)$,1)–FIGARCH(1,$d(v)$,1) model for each stock’s returns independently. This specification extends traditional GARCH by introducing fractional integration in both the mean (ARFIMA) and variance (FIGARCH) processes, capturing potential long-memory behavior in volatility. Consequently, large shocks can have persistent effects, while the fractionally integrated structure can smooth out some short-term noise and better accommodate volatility clustering. However, because the model is univariate, it cannot directly incorporate systemic market influences—i.e., it does not explicitly account for simultaneous volatility shocks across multiple assets.

\noindent The ARFIMA(1, \(d(m)\), 1) process with log returns \(x_{t}\) is described as:

\medskip

\[
\phi(L)\,(1 - L)^{d(m)} x_{t} \;=\; \theta(L)\,\varepsilon_{t},
\]

\medskip

\noindent where \(d(m)\) represents the fractional differencing parameter governing long memory in the mean, and \(\varepsilon_{t}\) is a white noise innovation. For volatility, the FIGARCH(1, \(d(v)\), 1) framework is described as:

\medskip

\[
\phi(L)\,(1 - L)^{d(v)}\,\varepsilon_{t}^{2} \;=\; \omega \;+\; [1 - \beta(L)]\,\nu_{t},
\]

\medskip

\noindent in which the parameter \(d(v)\) measures the degree of long-memory dependence in volatility and $\nu_{t}$ is the innovations term.

\noindent In contrast, the MHN–GARCH(1,1) approach exploits a common market factor and asset-specific parameters within a joint, multivariate estimation. Each asset’s variance can be influenced by the index variance as well as its own idiosyncratic dynamics. This structure better captures correlations among assets, especially in turbulent periods, since a shock in the index can simultaneously raise the conditional volatility of all stocks linked to that index. Consequently, the MHN–GARCH estimates may exhibit more coordinated changes across different stocks relative to the univariate ARFIMA–FIGARCH model. However, the model is more parameter-intensive and presupposes a factor structure (often with the index as the main risk driver), which increases estimation complexity and can introduce biases if the chosen factor specification is misspecified.

\noindent Empirically, we see that historical volatility often shows abrupt jumps, reflecting narrow-window price swings. The univariate ARFIMA–FIGARCH estimates generally smooth out short-run fluctuations and capture potential long-memory dynamics, while still responding to recent large shocks. Meanwhile, MHN–GARCH estimates sometimes closely track the ARFIMA–FIGARCH line but diverge, especially during market-wide volatility episodes, if the index factor exerts a strong influence on each stock’s variance. These observations underscore the importance of model choice. While historical volatility provides an empirical benchmark, ARFIMA–FIGARCH and MHN–GARCH offer forward-looking structures, with the latter also exploiting cross-asset interactions. Such differences highlight how modeling assumptions can meaningfully alter volatility estimates and their responsiveness to market shocks.

\noindent To account for time-varying expected returns, we specify a risk premium component that links the conditional mean of returns to the asset's own conditional variance. This follows the spirit of the GARCH-in-mean framework and reflects empirical findings that investors demand higher expected returns in periods of greater risk. This linkage also plays a key role in the dynamic asset allocation application developed later, where filtered conditional variances feed into optimal portfolio weights under CRRA utility.

\noindent In conclusion, the MHN–GARCH model provides a unified framework for modeling asset return dynamics in a multivariate, non-Gaussian setting. Each stock–index pair is modeled as a bivariate system, capturing both idiosyncratic and systematic volatility behavior. The affine structure and NIG innovations allow us to generate filtered paths of variance and correlation that are used in downstream applications including option pricing and intertemporal portfolio optimization. The structure is designed to be both tractable and flexible, allowing the model to be estimated efficiently using maximum likelihood and implemented in forward-looking simulations for risk and return evaluation.

\newpage

\subsection{MHN-GARCH(1,1) Parameter Estimates}

\noindent In this section, we estimate the MHN–GARCH model using maximum likelihood across each stock–index pair in the DJIA. The estimated parameters define the dynamic structure of conditional variances, covariances, and expected returns, which form the foundation for all forward-looking applications in this paper. These include dynamic portfolio choice, implied volatility surface construction, and simulations of wealth evolution. Parameters such as volatility persistence ($\beta$), risk sensitivity ($\lambda$), and cross-market correlation ($a$, $\lambda_{1,2}$) allow us to quantify how information propagates across assets and time. Understanding these dynamics is critical for both pricing and control decisions in environments where volatility and correlation evolve unpredictably.

\noindent While standard GARCH estimation typically focuses on univariate dynamics, our framework allows for the propagation of index shocks into asset-level variances and expected returns. This makes the estimation process more informative and delicate, as identification must balance empirical accuracy with theoretical tractability. We impose parameter bounds consistent with stationarity and positivity constraints to ensure stability, initializing estimates using sample moments. The use of NIG-distributed standardized residuals requires appropriate Jacobian adjustments within the log-likelihood function, reflecting the nonlinear transformation between returns and innovations.

\noindent The estimated parameters are economically interpretable and form the foundation for our subsequent option pricing and portfolio construction applications. For example, the estimated $\lambda$ and $\lambda_{1,2}$ parameters quantify the strength of systematic risk pricing. In contrast, the correlation-driver parameter--$a$--measures the speed and stability of volatility co-movement between assets and the index. These parameters collectively define the conditional variance–covariance structure and expected return process in each period, enabling forward-looking risk control and dynamic asset allocation.

\noindent Relative to alternative specifications, such as univariate ARFIMA and FIGARCH filters, our estimates reflect richer volatility dynamics with observable cross-asset interactions. As highlighted in prior work by \citet{Anderson2007} and \citet{ChristoffersenHestonJacobs2013}, capturing the joint evolution of risk factors in a multivariate setting can improve both the statistical fit and the economic relevance of GARCH-style models. The ability to model tail risk and correlation dynamics in tandem further supports using our structure in institutional decision-making contexts.

\noindent Table\ref{tab:combined-parameters} reports the parameter estimates of both MHN-GARCH (1,1) and ARFIMA (1,$d(m)$,1)-FIGARCH (1,$d(v)$,1)
estimated by maximum likelihood estimation\footnote{We maximize the log-likelihood with NIG probability density function to estimate the set of 7 parameters in the MHN-GARCH model: $\Theta = (\omega,\;\alpha,\;\beta,\;\theta,\;\lambda,\;a,\;\lambda_{1,2})$. The log-likelihood specification is -- $\ell(\Theta)=
\sum_{t=1}^{T}\!\Bigl[
\tfrac12\bigl\{\log h_{1,t}+\log h_{2,t}\bigr\}
-\log f(z_{1,t})-\log f(z_{2,t})
\Bigr]$}. The four parameters shared by both the MHN–GARCH and ARFIMA–FIGARCH specifications—mean return ($\mu$), baseline volatility ($\omega$), sensitivity ($\alpha$), and persistence
($\beta$)—exhibit notable differences when estimated under the multivariate versus univariate frameworks. Estimation was conducted using maximum likelihood with numerical optimization initialized by moment-based parameter guesses. To ensure stability, constraints were imposed on $\beta$ and $\alpha$ to preserve positivity and covariance stationarity. Particular attention was given to the $\lambda_{1,2}$ parameter, which can become weakly identified when the stock–index correlation is low. In such cases, convergence was improved by narrowing the parameter bounds based on observed correlation levels. We verified local convexity of the log-likelihood around the estimated optimum for each bivariate pair. Estimation was performed using a filtered likelihood approach with robust computation of the Jacobian for the NIG innovations

\noindent Although twelve structural parameters arise in the unrestricted bivariate system, the first set of five parameters, namely, $\omega_{1},\;\alpha_{1},\;\beta_{1},\;\theta_{1},\;\text{and},\;\lambda_{1}$ are index specific parameters and don't include the identification of time-varying correlation/correlation driver parameter $a$. Hence, the first row of Table~\ref{tab:combined-parameters} reports these parameters for the index DJI.  Accordingly, we retain the seven parameters most sensitive to joint volatility dynamics for all the remaining 30 assets in the portfolio and calculate the mean following the method aforementioned. This dimensionality reduction is not feasible in the ARFIMA–FIGARCH model given its univariate identification with no correlation driver, underscoring the informational gain from exploiting cross‐asset covariance.

\noindent The parameters $\lambda$ and $\lambda_{1,2}$ govern the sensitivity of conditional expected returns to filtered variance components. Specifically, $\lambda$ represents the market price of index risk and directly enters the return equation of the index, while$\lambda_{1,2}$ captures how each stock’s return responds to index-level volatility shocks. These parameters are central in both CRRA utility optimization and option pricing. In portfolio choice, they determine how the investor reallocates toward or away from stocks based on changing risk premia. In option pricing, they influence the skew and shape of the implied volatility surface through their effect on forward-looking return asymmetry and conditional variance.

\begin{sidewaystable}[htbp]
    \centering
    \resizebox{\textwidth}{!}{%
\begin{tabular}{ccccccccccccccccc}
    \toprule \\ \\
         & \multicolumn{8}{c}{\textbf{MHN-GARCH(1,1)}} 
         & \multicolumn{8}{c}{\textbf{ARFIMA(1,d(m),1)-FIGARCH(1,d(v),1)}} \\ \\
       Asset & $\boldsymbol\mu \times(E-05)$& $\boldsymbol{\omega}$ & $\boldsymbol{\alpha}$ & $\boldsymbol{\beta}$ & $\boldsymbol{\theta}$ & $\boldsymbol{\lambda}$ & $\boldsymbol{a}$ & $\boldsymbol{\lambda_{12}}$ & $\boldsymbol\mu$ & $\textbf{AR}(1)$ & $\textbf{MA}(1)$ & \textbf{d(m)} & $\boldsymbol\omega$ & $\boldsymbol\alpha$ & $\boldsymbol\beta$ & \textbf{d(v)} \\ \\
              & (Mean Return) & (Baseline Vol)  & (Sensitivity) & (Persistence) & (Risk-Premium/Leverage) & (Market Price of Risk) & (Correlation Driver) & (Co-Market Price of Risk) & (Mean Return)  &  &  & (Long-Memory in Mean) & (Baseline Vol) & (Sensitivity) & (Persistence) & (Long-Memory in Volatility) \\ \\
\cmidrule(lr){2-9} \cmidrule(lr){10-17} \\ \\

 DJI & 8.66 & 0.000001 & 0.000003 & 0.840298 & 0.151432 & 1.0 & - & - & 0.000481 & -0.942853 & 0.952319 & 0.000001 & 0.0 & 0.08949 & 0.966384 & 1.0 \\ \\
 MMM & 1.49 & 0.000009 & 0.0 & 0.840295 & 0.018549 & -0.390745 & 0.952087 & 0.000918 & -0.000129 & 0.477653 & -0.588525 & 0.08303 & 0.0 & 0.058279 & 0.998908 & 0.966412 \\ \\
 AXP & 0.0001 & 0.000006 & 0.000001 & 0.870202 & 0.230257 & 1.0 & 1.415957 & 0.108178 & 0.001248 & 0.361308 & -0.341408 & 0.000001 & 0.0 & 0.111411 & 0.985444 & 0.999999 \\ \\
 AMGN & 0.00014 & 0.000012 & 0.000001 & 0.757549 & -0.00916 & 0.96364 & 0.632954 & 0.004158 & 0.000007 & -0.227011 & 0.379922 & 0.0 & 0.0 & 0.052451 & 0.963875 & 0.66662 \\ \\
 AAPL & 0.00013 & 0.00002 & 0.000004 & 0.662254 & 0.171092 & 1.0 & 1.182302 & 0.179858 & 0.000886 & 0.697512 & -0.736491 & 0.024295 & 0.0 & 0.062415 & 0.898851 & 0.396723 \\ \\
 BA & -7.76 & 0.000048 & 0.000007 & 0.62976 & 0.058465 & -0.621945 & 1.36962 & 0.034394 & -0.000441 & -0.327999 & 0.290839 & 0.013832 & 0.000035 & 0.347057 & 0.591053 & 0.297119 \\ \\
 CAT & 0.00011 & 0.000023 & 0.000002 & 0.648263 & 0.001017 & 1.0 & 1.262953 & 0.026813 & 0.000813 & -0.348645 & 0.416858 & 0.0 & 0.00002 & 0.888581 & 0.91286 & 0.077704 \\ \\
 CVX & 0.00014 & 0.000018 & 0.000005 & 0.721096 & -0.005934 & 1.0 & 0.843095 & 0.055101 & 0.000463 & -0.214037 & 0.245531 & 0.0 & 0.0 & 0.014111 & 0.972695 & 1.0 \\ \\
 CSCO & 8.57 & 0.000012 & 0.0 & 0.701285 & 0.051159 & 1.0 & 0.9573 & 0.083617 & 0.000794 & -0.54988 & 0.592456 & 0.0 & 0.0 & 0.216691 & 1.0 & 1.0 \\ \\ 
 KO & 8.23  & 0.000004 & 0.000001 & 0.76784 & 0.0283 & 1.0 & 0.573631 & 0.03463 & 0.000223 & -0.319513 & 0.284481 & 0.014447 & 0.000007 & 0.223203 & 0.497908 & 0.320346 \\ \\
 DOW & -2.45 & 0.000019 & 0.000004 & 0.678403 & 0.077407 & -0.765006 & 1.089206 & 0.011707 & -0.000237 & -0.79376 & 0.811395 & 0.0 & 0.0 & 0.068869 & 0.984698 & 0.999997 \\ \\
 GS & 0.00011 & 0.000018 & 0.000002 & 0.644568 & 0.067112 & 1.0 & 1.284256 & 0.085429 & 0.001006 & -0.116206 & 0.198657 & 0.000047 & 0.0 & 0.333795 & 0.982525 & 0.920361 \\ \\
 HD & 0.00013 & 0.000012 & 0.000005 & 0.698637 & 0.035517 & 1.0 & 1.112192 & 0.091865 & 0.000579 & -0.376145 & 0.4181 & 0.000001 & 0.0 & 0.014489 & 0.973496 & 0.965846 \\ \\
 HON & 6.42 & 0.000009 & 0.0 & 0.68818 & -0.014114 & 0.674985 & 1.032844 & -0.032961 & 0.000238 & -0.746973 & 0.789089 & 0.0 & 0.0 & 0.242891 & 0.975243 & 0.871838 \\ \\
 IBM & 8.83 & 0.000011 & 0.000001 & 0.725614 & 0.010899 & 1.0 & 0.817048 & 0.073229 & 0.000608 & -0.664645 & 0.689875 & 0.0 & 0.000001 & 0.003528 & 0.900729 & 0.404626 \\ \\
 INTC & -0.00013 & 0.00004 & 0.000013 & 0.638717 & -0.104405 & -0.906747 & 1.317985 & -0.019525 & -0.000398 & 0.205172 & -0.096528 & 0.016669 & 0.000001 & 0.248548 & 0.987716 & 0.937291 \\ \\
 JNJ & 6.55 & 0.000007 & 0.000001 & 0.785671 & -0.029059 & 0.683787 & 0.470837 & -0.04355 & 0.000086 & 0.973238 & -0.988951 & 0.010186 & 0.000001 & 0.201211 & 0.947059 & 0.800581 \\ \\
 JPM & 0.00011 & 0.000013 & 0.000002 & 0.674442 & 0.172963 & 1.0 & 1.112038 & 0.170075 & 0.000917 & 0.443154 & -0.398278 & 0.000001 & 0.000001 & 0.062567 & 0.973731 & 0.999985 \\ \\
 MCD & 9.11 & 0.000002 & 0.000001 & 0.883167 & 0.058064 & 1.0 & 0.749812 & 0.047682 & -0.009716 & -0.119835 & -0.634712 & 0.5 & 0.000003 & 0.632217 & 0.899504 & 0.026901 \\ \\
 MRK & 0.0001 & 0.000011 & 0.000001 & 0.799206 & -0.017951 & 0.975698 & 0.392815 & -0.012797 & 0.000085 & -0.849216 & 0.828701 & 0.000005 & 0.0 & 0.132643 & 1.0 & 0.999994 \\ \\
 MSFT & 0.00013 & 0.000004 & 0.000001 & 0.921528 & 0.132346 & 1.0 & 1.149243 & 0.102366 & 0.000934 & 0.524142 & -0.583648 & 0.002687 & 0.000001 & 0.030844 & 0.969095 & 0.983855 \\ \\ 
 NKE & -5.11 & 0.000035 & 0.0 & 0.655882 & 0.043161 & -0.89331 & 1.219035 & 0.026398 & -0.000565 & 0.85718 & -0.877144 & 0.000368 & 0.000001 & 0.055519 & 0.99206 & 0.936636 \\ \\
 NVDA & 0.0004 & 0.000145 & 0.000012 & 0.561716 & 0.297195 & 1.0 & 1.761874 & 0.192384 & 0.002943 & -0.438235 & 0.402254 & 0.001274 & 0.000002 & 0.000006 & 0.980341 & 0.999999 \\ \\
 PG & 0.0001 & 0.000002 & 0.000001 & 0.917734 & 0.277557 & 1.0 & 0.539736 & -0.052227 & 0.000204 & 0.577856 & -0.647021 & 0.043107 & 0.0 & 0.047167 & 0.898113 & 0.393587 \\ \\
 CRM & 0.00012 & 0.000038 & 0.000006 & 0.653484 & 0.097962 & 0.537626 & 1.232857 & 0.119671 & 0.000374 & -0.737859 & 0.789322 & 0.01307 & 0.000004 & 0.332544 & 0.949829 & 0.773336 \\ \\
 TRV & 0.0001 & 0.000013 & 0.000001 & 0.724498 & 0.023047 & 1.0 & 0.823479 & 0.019258 & 0.000658 & 0.498054 & -0.586675 & 0.000003 & 0.0 & 0.007997 & 1.0 & 0.999999 \\ \\
 UNH & 0.0001 & 0.000011 & 0.000003 & 0.737053 & 0.048746 & 1 & 0.751106 & 0.035184 & 0.000542 & -0.544061 & 0.531575 & 0.0 & 0.0 & 0.034612 & 1.0 & 1.0 \\ \\
 VZ & -3.577 & 0.000007 & 0.000003 & 0.788852 & 0.007576 & -0.980932 & 0.452501 & -0.01252 & -0.000283 & -0.687663 & 0.676956 & 0.0 & 0.00002 & 0.539728 & 0.608776 & 0.19817 \\ \\
 V & 8.183 & 0.00001 & 0.000002 & 0.691946 & 0.080326 & 1.0 & 1.011134 & 0.112453 & 0.000302 & -0.42155 & 0.43566 & 0.008878 & 0.000001 & 0.022612 & 0.89849 & 0.400271 \\ \\
 WBA & -0.00015 & 0.000031 & 0.000007 & 0.694729 & -0.060096 & -1.0 & 0.995089 & -0.070209 & -0.001089 & 0.89598 & -0.911379 & 0.041802 & 0.000001 & 0.046237 & 0.986938 & 0.966851 \\ \\
 WMT & 8.652 & 0.000008 & 0.000001 & 0.777441 & 0.041832 & 1.0 & 0.518281 & 0.053403 & 0.000274 & -0.546409 & 0.544064 & 0.013396 & 0.000001 & 0.163915 & 0.963076 & 0.884018 \\ \\
\bottomrule
\end{tabular}
    }
     \caption{\centering Parameter Estimates. The table on the left reports parameter estimates of the MHN-GARCH (1,1) specification, and the table on the right reports parameter estimates of the ARFIMA (1, $d(m)$,1)-FIGARCH (1,$d(v)$,1) specifications. Both structural models have been estimated by MLE. Common parameters include $\mu$, $\omega$, $\alpha$, and $\beta$. Newly identified parameters include $\theta$, $\lambda$, $a$, and $\lambda_{1,2}$. In the ARFIMA-FIGARCH specification, we also report the additional parameters nested in the model, i.e., AR(1), MA(1), $d(m)$, and $d(v)$.}
    \label{tab:combined-parameters}
\end{sidewaystable}

\noindent Table~\ref{tab:combined-parameters} reports the parameter estimates of both MHN-GARCH (1,1) and ARFIMA (1,$d(m)$,1)-FIGARCH (1,$d(v)$,1) estimated by maximum likelihood estimation. The four parameters shared by both the MHN–GARCH and ARFIMA–FIGARCH specifications—mean return ($\mu$), baseline volatility ($\omega$), sensitivity ($\alpha$), and persistence ($\beta$)—exhibit notable differences when estimated under the multivariate versus univariate frameworks.  To illustrate these contrasts, we focus on the highest, median, and lowest estimates across our 31‐asset universe. All returns are measured in continuously compounded units and rescaled into percentages for interpretation of these estimates.

\noindent Under MHN–GARCH, McDonald’s (MCD) displays the largest mean return at \(0.00911\%\), whereas Boeing (BA) has the most negative estimate at \(-0.00776\%\).  Apple (AAPL) sits near the median with \(0.00130\%\).  In the ARFIMA–FIGARCH model, MCD’s mean return is slightly negative at \(-0.9716\%\), reflecting long–memory adjustment, while BA remains negative but smaller in magnitude at \(-0.0441\%\).  AAPL’s ARFIMA mean of \(0.0886\%\) exceeds its MHN–GARCH counterpart by nearly an order of magnitude, indicating that univariate long–memory filtering can materially alter expected‐return estimates when systemic correlations are ignored.

\noindent For baseline volatility, NVIDIA (NVDA) registers the highest short-run variance parameter of \(0.0145\%\), capturing its pronounced intraday swings, whereas the DJIA index has the lowest at \(0.00010\%\).  AAPL again is near the median at \(0.00200\%\).  Under ARFIMA–FIGARCH, many assets have effectively zero baseline volatility—reflecting that the FIGARCH long–memory component absorbs most of the dynamics—so AAPL and DJI both report \(0\%\).  Thus, MHN–GARCH attributes more fluctuation to the immediate‐reaction term than does the univariate FIGARCH.

\noindent The sensitivity coefficient \(\alpha\) is highest for NVDA at \(0.00120\%\), while CSCO and MMM have \(0\%\), indicating negligible short‐run response.  AAPL’s sensitivity of \(0.00040\%\) again approximates the cross‐sectional median.  In ARFIMA–FIGARCH, NVDA’s \(\alpha\) is \(0.00060\%\), and most other assets—including AAPL and CSCO—also report zero, suggesting heavy‐tailed innovations supplant fast‐moving variance effects.

\noindent Persistence (\(\beta\)) under MHN–GARCH peaks at MSFT with \(92.15\%\) and is lowest for NVDA at \(56.17\%\); AAPL’s persistence of \(66.23\%\) is mid‐sample.  In ARFIMA–FIGARCH, MSFT’s \(\beta\) rises to \(96.91\%\) and NVDA’s to \(98.03\%\), reflecting that the FIGARCH long–memory term further elevates persistence.  These contrasts underscore how univariate long‐memory models allocate more variance persistence than the multivariate MHN–GARCH filter.

\noindent Turning to MHN–GARCH–specific parameters, NVDA shows the largest risk premium \(\theta\) of \(29.72\%\), while AMGN has the smallest at \(-0.92\%\); AAPL’s median risk premium is \(17.11\%\).  The market‐price‐of‐risk \(\lambda\) is unity for most stocks, with VZ at \(-98.09\%\) and HON at \(67.50\%\) illustrating the cross‐sectional spread.  The correlation‐driver \(a\) peaks at \(176.19\%\) for NVDA and is lowest at \(57.36\%\) for KO, with GS’s median loading at \(128.43\%\).  Finally, the co‐market price of risk \(\lambda_{12}\) ranges from \(-3.30\%\) for HON up to \(19.24\%\) for NVDA (median: \(8.54\%\) for GS), highlighting substantial heterogeneity in how index shocks transmit into each asset’s expected return.  

\noindent While both the MHN–GARCH and ARFIMA–FIGARCH models achieve good statistical fit, their economic implications differ meaningfully. The MHN–GARCH model captures time-varying covariances and systematic risk pricing, allowing filtered returns to reflect market-wide volatility conditions. In contrast, the ARFIMA–FIGARCH model treats each asset in isolation, modeling long memory in volatility but ignoring co-movement and shock transmission. This leads to structurally flatter implied volatility surfaces and underestimates of systemic exposure. The ability of MHN–GARCH to incorporate dynamic correlation and non-Gaussian innovations makes it more suitable for portfolio control and derivative pricing under realistic market stress scenarios, consistent with findings in \citet{Anderson2007} and \citet{ChristoffersenHestonJacobs2013}.

\subsection{Option Pricing} \label{options}

\noindent We next apply the estimated MHN–GARCH parameters to price European options using a closed-form expression derived from \citet{hn2000}. While the original HN framework applies to univariate GARCH processes, we extend this approach by integrating filtered volatilities from our bivariate stock–index specification. This extension enables us to account for joint tail risk and index-driven volatility clustering—features commonly observed during market stress but often omitted in traditional option pricing models. The resulting pricing formula incorporates asset-specific variance, cross-asset dependence, and filtered innovations, allowing us to derive realistic implied volatility surfaces across a wide range of strikes and maturities.

\noindent Unlike the classical Black–Scholes formula, which assumes constant volatility and log-normal returns, our option valuation incorporates filtered conditional variance, which evolves dynamically over time and reflects the joint risk structure between the asset and the market index. This approach allows us to price options in a setting where downside risk, volatility asymmetry, and index contagion are endogenously captured. It also preserves analytical tractability, enabling fast evaluation of option prices across portfolios with heterogeneous risk exposures.

 \noindent Consider the characteristic function of the log spot price, $f(i\phi)$ in \citet{hn2000}, then

\begin{align}
    E_{t}[\text{Max}(S(T) - K,0)] &= f(1)\left(\frac{1}{2} + \frac{1}{\pi}\int_{0}^{\infty}\text{Re}\left[\frac{K^{-i\phi}f(i\phi + 1)}{i\phi f(1)}\right]d\phi\right) - \notag \\[1mm]
    & K\left(\frac{1}{2} + \frac{1}{\pi}\int_{0}^{\infty}\text{Re}\left[\frac{K^{-i\phi}f(i\phi)}{i\phi}\right]d\phi\right),
\end{align}

\noindent where Re$[\cdot]$ indicates the real part of the complex number. Based on Eq. 7, an option value is the discounted expected value of the payoff payoff, $\text{Max(S(T)-K,0)}$, calculated using the risk-neutral probabilities which is enabled by the use of the characteristic value of the function, $f^{*}(i\phi)$. Therefore, the European option value with strike price $K$ with expiration $T$ at time $t$ is defined as

\begin{align}
    C &= e^{-r(T-t)}E_{t}^{*}[\text{Max}(S(T)-K,0)] \notag \\[1mm]
    &= \frac{1}{2}S(t) + \frac{e^{-r(T-t)}}{\pi}\left(\frac{1}{2} + \frac{1}{\pi}\int_{0}^{\infty}\text{Re}\left[\frac{K^{-i\phi}f(i\phi + 1)}{i\phi f(1)}\right]d\phi\right) \notag \\[1mm]
    & - K e^{-r(T-t)}\left(\frac{1}{2} + \frac{1}{\pi}\int_{0}^{\infty}\text{Re}\left[\frac{K^{-i\phi}f(i\phi)}{i\phi}\right]d\phi\right)
\end{align}

\noindent where $E_{t}^{*}$ is the expectation of the option value priced under the risk-neutral distribution. As described in the original closed-form solution, this valuation formula, contrary to the classical Black-Scholes formula for option valuation, is a function of the current asset price, $S(t)$, and the conditional variance, $h_{i,t}$. Given that $h_{i,t}$ is a function of the observed path of the asset price, the extended option pricing formula is a function of the current and lagged asset prices. With the identification of conditional volatility from the joint probability distribution of the index and asset pair in the bivariate strategy, the variance process enters the option valuation framework with cross-asset price risk factored into the filtered volatility, hence, there is no need for separate identification using the correlation driver for the option valuation formula to construct call and put price implied volatility surfaces.

\begin{figure}[htbp]
    \centering
    \includegraphics[width=\textwidth,%
      height=\textheight,%
      keepaspectratio]{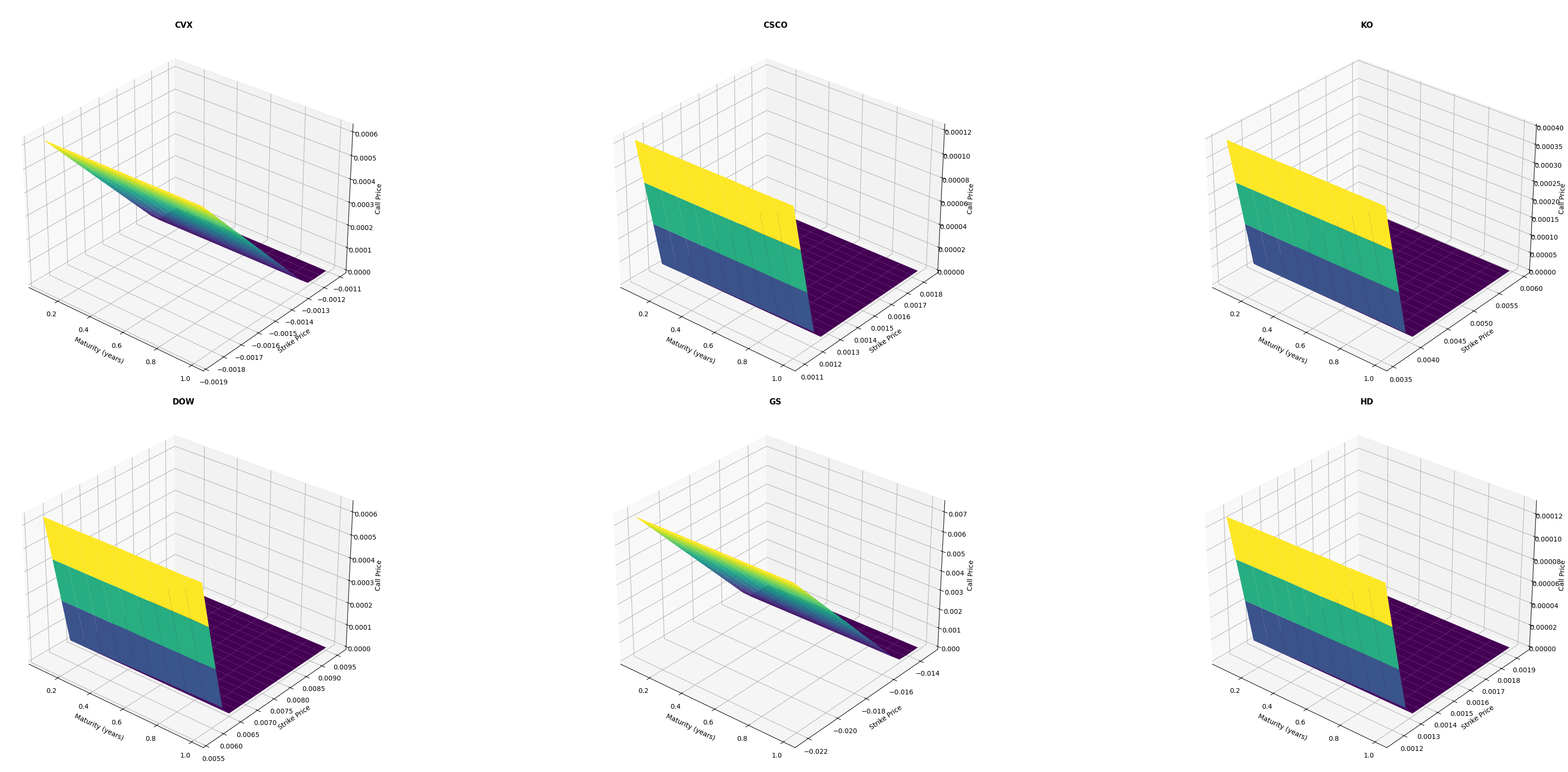}
    \caption{Call Option}
    \label{fig:call_1}
\end{figure}

\begin{figure}[htbp]
    \centering
    \includegraphics[width=\textwidth,%
      height=\textheight,%
      keepaspectratio]{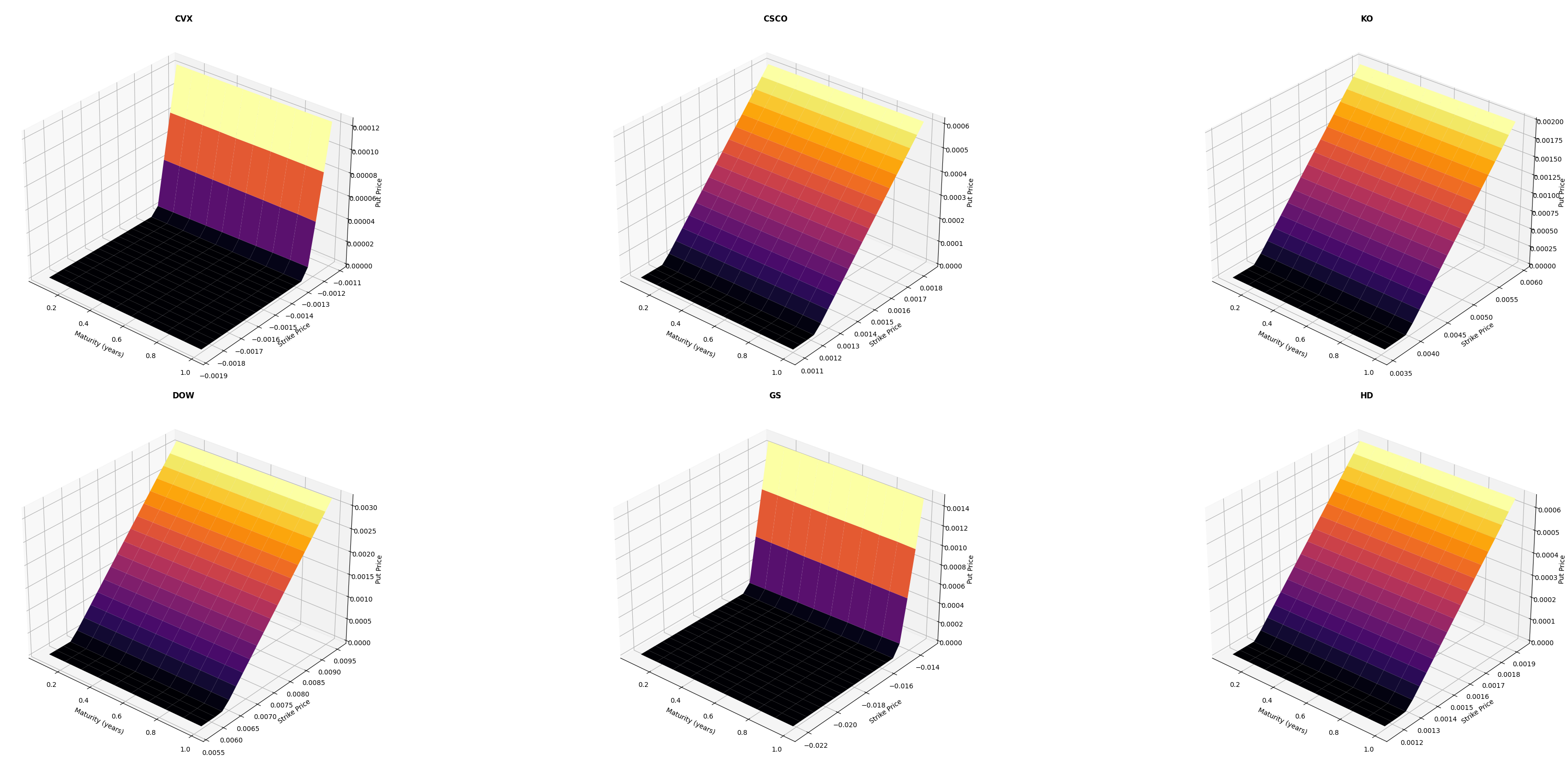}
    \caption{Put Option}
    \label{fig:put_1}
\end{figure}

\noindent Figures ~\ref{fig:call_1} and \ref{fig:put_1} present the implied volatility surfaces for selected DJIA constituents, constructed from our MHN–GARCH calibrated option pricing model. These surfaces exhibit notable patterns: downward sloping skews in call options, pronounced smirks in puts, and maturity-dependent curvature—all of which mirror empirical findings in equity option markets. These features validate the model's ability to capture asymmetric return distributions, heavy tails, and volatility clustering consistent with documented option market anomalies.

\noindent In our MHN–GARCH surfaces, a downward‐sloping skew in call options—where implied volatility is highest for low strikes and falls as strikes rise—reflects traders’ willingness to pay a premium for protection against large upward jumps (e.g., takeover bids or unexpected positive news), while a pronounced smirk in puts—where volatility steeply increases for strikes below the spot—signals heightened demand for insurance against extreme downside moves (such as crashes or negative earnings shocks).  The maturity‐dependent curvature, with sharper bends in near‐term slices and flatter profiles at longer horizons, captures the concentration of short‐run event risk around known dates (earnings, macro releases) versus a more averaged view of uncertainty over extended periods, illustrating how option‐implied volatilities encode both strike‐specific asymmetries and the term structure of market risk.

\noindent These differences in implied volatility profiles highlight the importance of modeling joint risk factors. For assets like CVX, where commodity price shocks drive extreme moves, the model's ability to embed tail risk leads to more convex pricing at low strikes, for more stable names like CSCO, flatter volatility surfaces result from weaker leverage effects and more symmetric innovations. This granularity is challenging to replicate using univariate GARCH models or simple historical estimates.

\noindent Turning to the put‐price implied volatility surfaces, KO and DOW reveal markedly convex profiles in strike, particularly for shorter maturities, producing the familiar “smirk” shape. The steep put prices at deep‐in‐the‐money levels reflect the negative skewness of filtered innovations—captured by the MHN‐GARCH’s asymmetric response to large negative returns—and the heightened marginal probability of adverse price moves in these more defensive or broad‐market assets. By contrast, CSCO and HD show less pronounced smirks, with put‐price surfaces that rise relatively linearly in strike. Given more symmetric innovation distributions, evident in the lower calibrated leverage coefficients, they produce muted tail‐risk premia and hence shallower implied‐volatility slopes.

\noindent Across all six assets, the maturity dimension further reveals the degree of volatility persistence: steeper curvature in the near‐term slice signals rapid mean‐reversion in the conditional variance, while flatter long‐term slopes point to convergence toward the unconditional variance. Our joint MHN‐GARCH framework allows this term‐structure behavior to vary by asset and in relation to the index, capturing interdependencies that univariate models cannot. In sum, these surfaces demonstrate that option prices—and the implied volatility smiles and smirks they produce—are materially shaped by multivariate risk dynamics. By filtering historical innovations through an affine multivariate GARCH approach, we present a comprehensive, data‐driven characterization of joint tail behavior, correlation shifts, and asymmetric responses, thereby advancing the accuracy and interpretability of option valuation in a multi‐asset framework.

\noindent The option pricing results reinforce the value of our joint volatility filtering approach. The model yields option values that align with market-observed behaviors by embedding time-varying volatility, asymmetric innovation distributions, and cross-asset correlation into the pricing kernel. These features are also critical in dynamic portfolio construction, as they provide forward-looking indicators of skewness, tail risk, and implied correlation that affect asset allocation decisions. In the next section, we show how these same filtered parameters shape the evolution of optimal wealth under CRRA utility.

\section{Results} \label{empirical}

\noindent We now turn to the empirical results derived from applying the MHN–GARCH(1,1) model to our 31-asset universe composed of the 30 DJIA constituents and the DJIA index itself. Having estimated the joint volatility dynamics and calibrated the option pricing model in the preceding sections, we evaluate the economic implications of the filtered parameters across three key applications: dynamic portfolio optimization under CRRA preferences, construction of mean–variance efficient frontiers, and calculation of wealth-equivalent losses (WEL) relative to static benchmarks. These analyses provide a comprehensive assessment of how the multivariate affine GARCH framework enhances investor decision-making under realistic market conditions, particularly by capturing heavy tails, evolving correlations, and asymmetric volatility propagation.

\noindent We begin in Section~\ref{ipc} by examining optimal wealth allocation strategies, highlighting how dynamic variance–covariance filtering translates into forward-looking portfolio weights and wealth allocation. This is followed in Section~\ref{ef} by a comparison of efficient frontiers under our proposed model and an alternative univariate FIGARCH benchmark. Section~\ref{wel} quantifies the utility loss from ignoring time-varying risks through the lens of wealth-equivalent losses. Together, these results establish the practical value of modeling joint volatility dynamics in financial decision-making.

\subsection{Optimal Wealth Allocation} \label{ipc}

\noindent We formulate the investor’s problem as a discrete-time dynamic programming problem, solving the Bellman equation to maximize terminal wealth under a constant relative risk aversion (CRRA) utility function. This optimization is conducted within a multivariate Heston–Nandi GARCH(1,1) framework, where conditional variances and covariances evolve dynamically based on filtered data. By solving for the optimal policy using backward induction, we derive a fully time-varying portfolio allocation rule that adjusts endogenously to evolving volatility and correlation forecasts.

\noindent This formulation is particularly relevant for both retail and institutional investors who routinely hedge individual stock positions with index exposure using bivariate trading strategies. In such settings, treating each asset’s variance in isolation overlooks cross-asset spillovers and dynamic co-movements that drive true portfolio risk. Our MHN–GARCH filter captures these features by allowing key parameters—volatility persistence, leverage, risk premia, and the market price of risk—to vary over time. Embedding these into the Bellman framework yields allocation paths that adapt dynamically to changes in market conditions, mitigating downside risk and enhancing long-run growth.

\begin{figure}[htbp]
    \centering
    \includegraphics[width=1\linewidth]{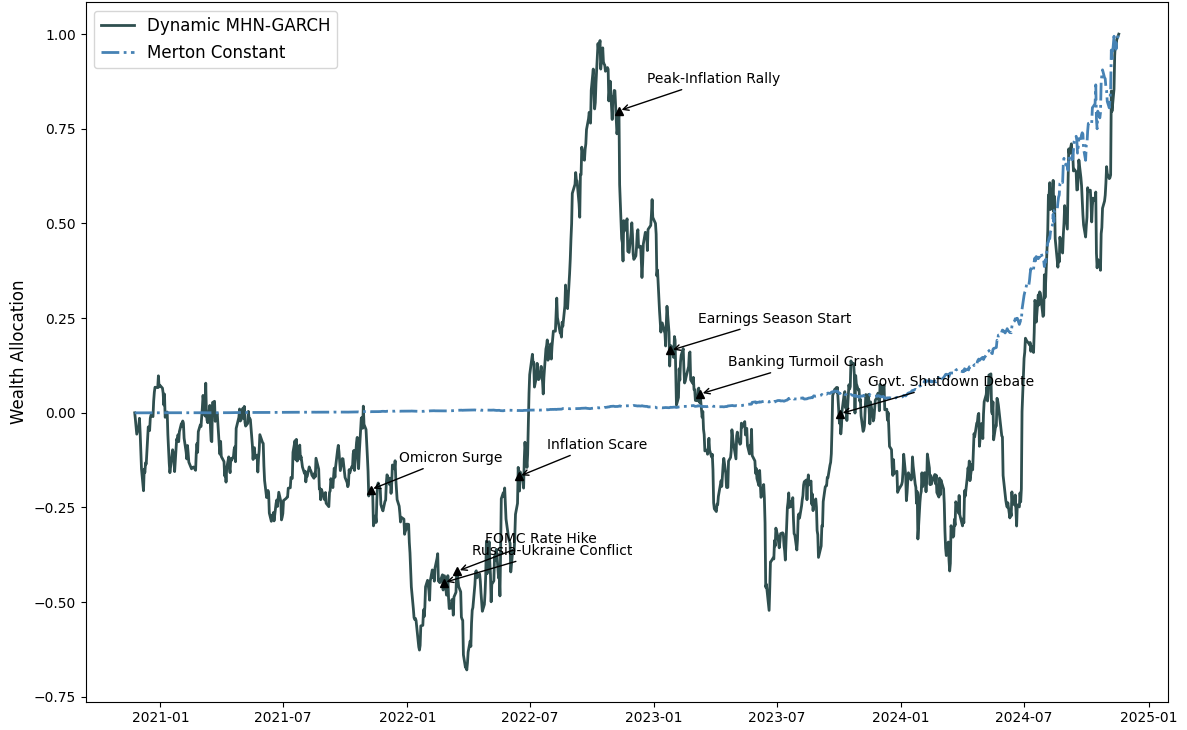}
    \caption{Wealth Allocation Dynamics of a Risk-Averse Investor}
    \label{fig:wealth_allocation}
\end{figure}

\noindent Figure \ref{fig:wealth_allocation} illustrates the evolution of a CRRA investor’s wealth allocation across time when guided by the MHN–GARCH filter. Several important insights emerge from this analysis, particularly with respect to adaptive risk management, precautionary saving, intertemporal risk sharing, and relative performance versus static benchmarks.

\noindent First, adaptive risk management is a core feature of this framework. By modeling conditional variance and correlation dynamically, the investor adjusts portfolio weights in response to short-term forecasts of increased volatility or adverse co-movement. Risky asset exposure is reduced during turbulent periods, limiting potential losses, and expanded during stable periods when expected returns are more favorable. This responsiveness contrasts sharply with the static Merton benchmark, which maintains fixed allocations irrespective of shifting risk.

\noindent Second, the CRRA utility function induces a form of precautionary saving behavior. This prudence leads the investor to tilt allocations toward safer assets when market volatility surges or asset correlations become less favorable. Such behavior provides a critical buffer against negative shocks and helps smooth marginal utility over time. For instance, during the “Omicron Surge” or the “Banking Turmoil Crash,” the dynamic strategy significantly curtails exposure to high-risk assets, thereby demonstrating a hedging motive.

\noindent Third, the Bellman optimization structure enables intertemporal risk sharing. The model distributes risk exposure over time in anticipation of future volatility and correlation states. When uncertainty rises, capital is reallocated toward lower-risk assets. As volatility subsides, the model redeploys capital into riskier holdings, producing a smoother, more strategic wealth accumulation over the investment horizon.

\noindent Lastly, comparing the MHN–GARCH strategy to a static constant-variance benchmark reveals meaningful performance differences. The static Merton allocation fails to respond to evolving market conditions, potentially exposing the investor to unanticipated shocks or missed opportunities. In contrast, the dynamic allocation strategy rebalances in a forward-looking fashion, capturing favorable risk-return tradeoffs as they emerge. This superior timing leads to a higher cumulative wealth path and a more efficient use of capital over time, as later substantiated by the efficient frontier and wealth-equivalent loss results in Sections 4.2 and 4.3.

\noindent The practical implications are clear: dynamic allocation under MHN–GARCH significantly reduces drawdowns in volatile environments while capturing upside during calmer regimes. Over the long term, this approach results in steadier wealth growth and enhanced investor utility. The observed wealth trajectories validate the theoretical advantage of incorporating joint filtered conditional volatility and correlation into portfolio optimization. A prudent CRRA investor who re-optimizes allocations based on these dynamics enjoys both improved risk mitigation and superior long-term performance.

\subsection{Efficient Frontier} \label{ef}

\noindent To evaluate the economic implications of our multivariate modeling approach, we construct Markowitz-style efficient frontiers (\citet{Markowitz1952} and \citet{Markowitz1959}) based on dynamic portfolio weights generated from the MHN–GARCH(1,1) model. This exercise quantifies the attainable risk–return trade-offs and enables direct comparison with an alternative univariate framework, specifically the ARFIMA–FIGARCH model calibrated with Normal Inverse Gaussian (NIG) innovations. By solving for the optimal portfolio at each rebalancing date, we assess how modeling joint time-varying volatility and heavy tails improves mean–variance efficiency in a realistic multi-asset setting.

\begin{figure}[h]
    \centering
    \begin{minipage}[t]{0.49\textwidth} 
        \centering
        \includegraphics[width=\textwidth]{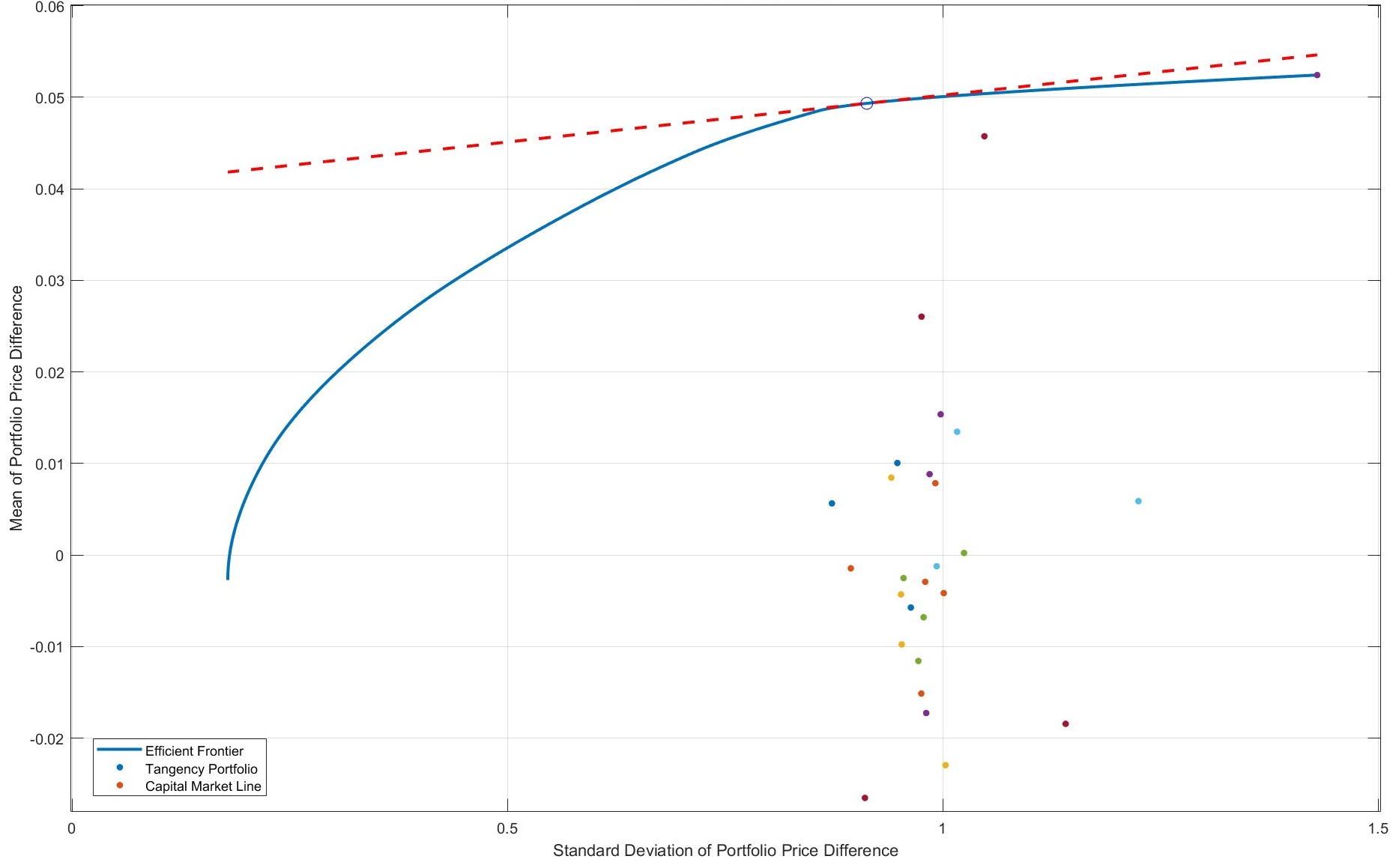} 
        \label{fig:markowitz_hist}
    \end{minipage}
    \hfill
    \begin{minipage}[t]{0.49\textwidth} 
        \centering
        \includegraphics[width=\textwidth]{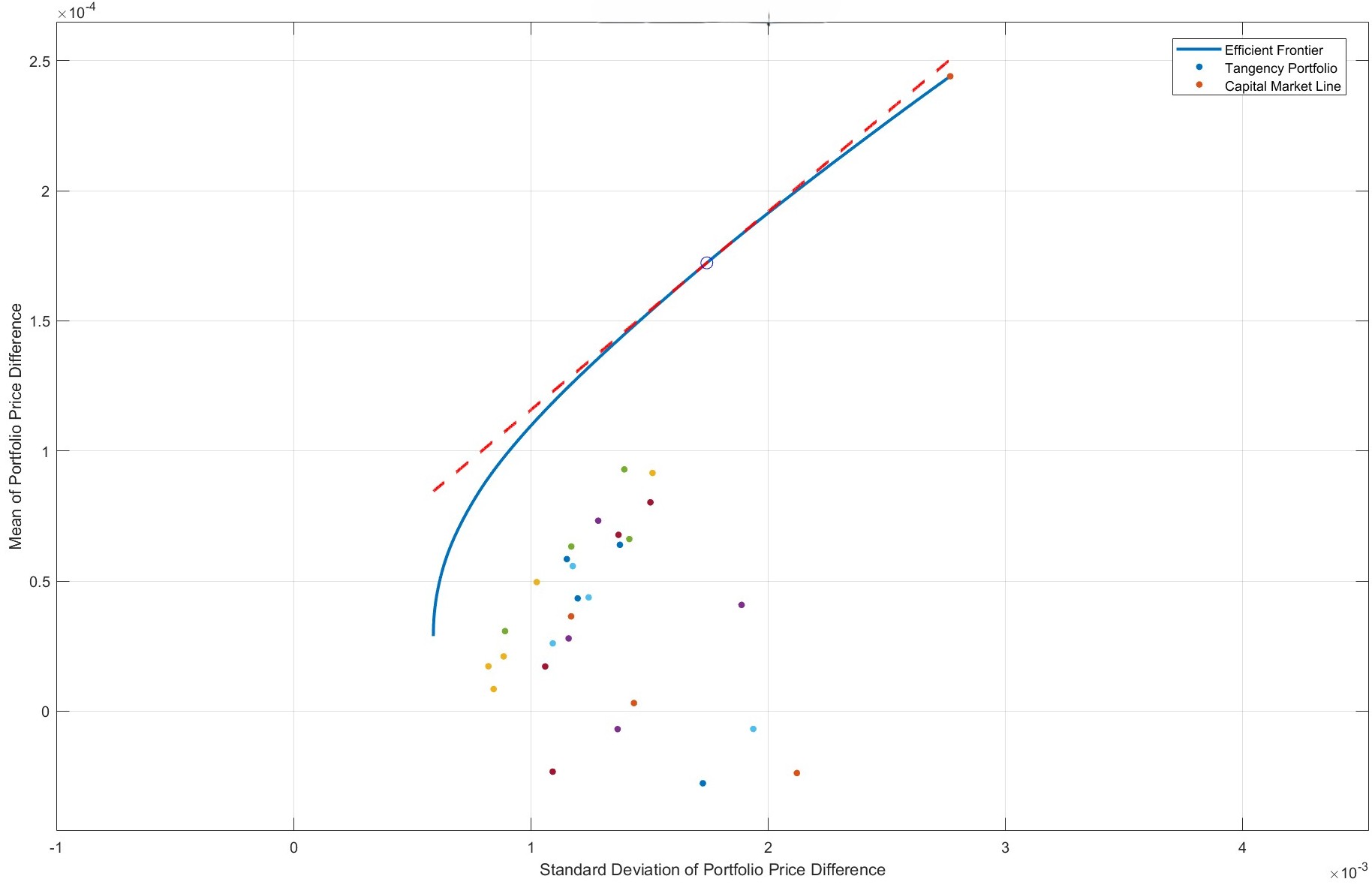} 
        \label{fig:markowitz_mhn}
    \end{minipage}
    \caption{Markowitz Optimization-based Efficient Frontier}
    \label{fig:Markowitz_EF}
\end{figure}

\noindent Figure~\ref{fig:Markowitz_EF} presents two efficient frontiers for the 31-asset universe (30 DJIA constituents and the DJIA index). The left panel corresponds to portfolios optimized using conditional variances from the univariate ARFIMA–FIGARCH model, while the right panel uses conditional covariance matrices generated by the joint MHN–GARCH filter. For both models, innovations are drawn 10,000 times from their estimated NIG distributions to simulate the impact of non-Gaussian risk. Dynamic tangency portfolios are plotted along each frontier, with corresponding Capital Market Lines (CMLs) overlaid in dashed red.

\noindent The results demonstrate that the MHN–GARCH model produces a systematically superior efficient frontier. At a target portfolio standard deviation of 0.75\%, the MHN–GARCH approach delivers a mean return of approximately 0.049\%, compared to 0.044\% under ARFIMA–FIGARCH. This difference, though seemingly small, represents a meaningful gain in mean–variance efficiency, especially when compounded over time. The improvement stems from the multivariate model’s ability to internalize dynamic correlations, leverage effects, and cross-asset volatility transmission—features that the univariate specification cannot fully accommodate.

\noindent The tangency portfolios further reinforce this distinction. Under MHN–GARCH, the tangency portfolio achieves a Sharpe ratio of approximately 0.175, substantially exceeding the 0.05 Sharpe ratio under ARFIMA–FIGARCH. This reflects the multivariate model’s superior ability to capture time-varying risk premia arising from joint index–asset interactions. Moreover, the steeper slope of the MHN–GARCH CML implies a more favorable reward-to-risk ratio for marginal increases in volatility—a critical insight for institutional investors managing capital across complex portfolios.

\noindent Importantly, these results are not limited to differences in marginal returns. The full distribution of returns is shaped differently under each model. ARFIMA–FIGARCH assumes a static covariance structure and isolates asset volatilities, potentially underestimating joint downside risks. In contrast, MHN–GARCH produces dynamically updated covariance matrices, leading to more robust diversification benefits and reduced exposure to extreme co-movement during market stress.

\noindent To our knowledge, this is the first implementation of a fully multivariate NIG-innovation-based efficient frontier using dynamically estimated conditional covariances. This approach bridges modern econometric filtering with classic portfolio theory, offering a forward-looking method for constructing optimal portfolios under realistic risk dynamics. The practical implication is clear: incorporating joint stochastic volatility and tail risk delivers superior mean–variance outcomes, validating the MHN–GARCH framework as a valuable tool for advanced asset allocation and risk budgeting.

\subsection{Wealth-Equivalent Loss Analysis} \label{wel}

\noindent To quantify the economic cost of using a suboptimal portfolio strategy, we compute Wealth-Equivalent Losses (WEL) comparing the dynamic MHN–GARCH-based allocation to a benchmark Merton constant-variance strategy. The WEL expresses the utility shortfall from adopting the benchmark as the percentage reduction in initial wealth that would make the investor indifferent between the two strategies. This transformation of utility differences into a monetary metric makes the welfare implications of time-varying risk models immediately interpretable.

\noindent The Merton solution serves as a natural benchmark due to its closed-form elegance under constant investment opportunities. However, it fails to incorporate time-varying volatility, cross-asset correlation, or heavy-tailed risk—features that are empirically prominent and economically significant. By comparing the MHN–GARCH strategy against this baseline, we directly assess the value added by modeling realistic risk dynamics.

\begin{figure}[htbp]
    \centering
    \includegraphics[width=\textwidth,%
      height=\textheight,%
      keepaspectratio]{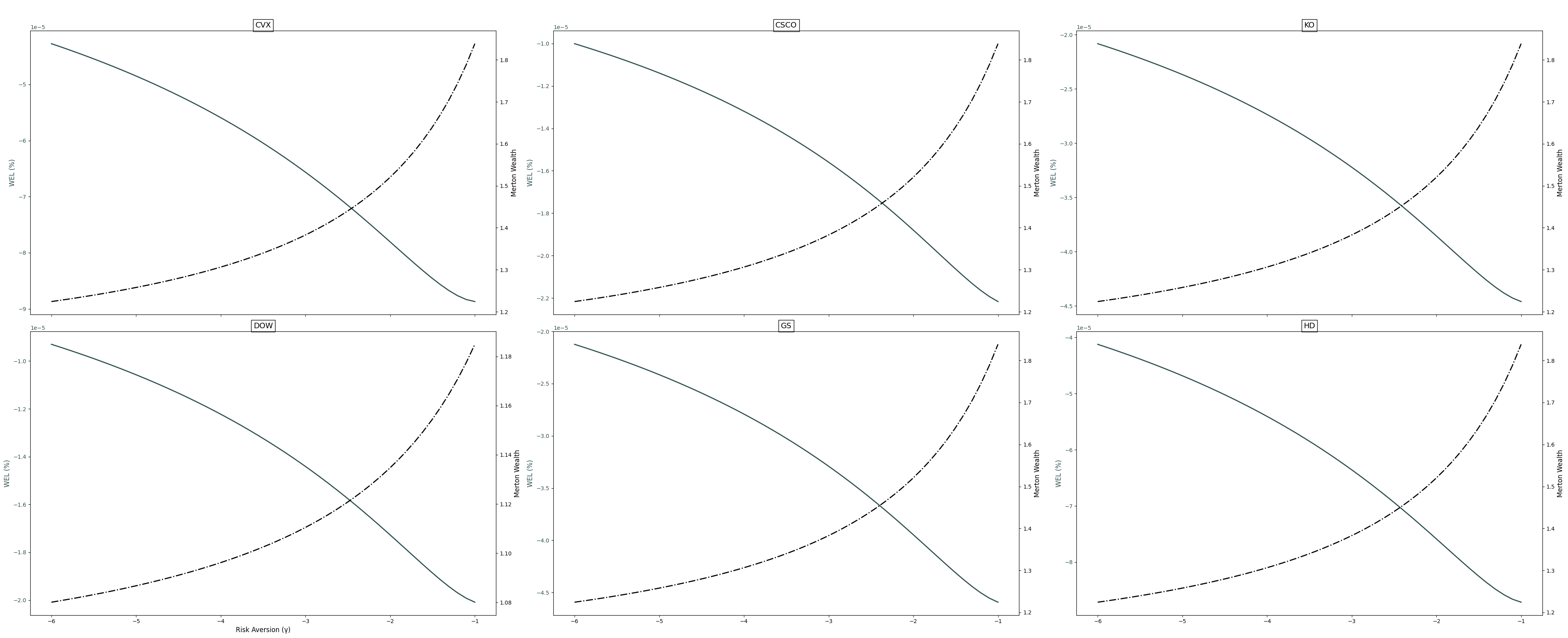}
    \caption{Wealth Equivalent Loss}
    \label{fig:wel_1}
\end{figure}

\noindent Figure \ref{fig:wel_1} plots WEL values for six DJIA constituents—CVX, CSCO, KO, DOW, GS, and HD—across a range of investor risk aversion coefficients (denoted $\gamma$ in the CRRA framework). In every case, the WEL is strictly negative, indicating that the MHN–GARCH strategy consistently delivers higher expected utility. Furthermore, as $\gamma$ increases (i.e., as the investor becomes more risk-averse), the magnitude of the WEL also increases, highlighting the growing importance of dynamic risk modeling for conservative investors.

\noindent For example, CVX exhibits the highest WEL gains: ranging from approximately $−0.0091\%$ at $\gamma = 1$ to $−0.0062\%$ at $\gamma = 6$. HD shows a similarly strong profile, with WEL ranging from $−0.0088\%$ to $−0.0052\%$. These results suggest that investors in sectors exposed to commodity shocks or construction-cycle volatility gain substantial utility from dynamically filtering and responding to market conditions. For GS, the WEL spans $−0.0046\%$ to $−0.0022\%$, reflecting the benefits of accounting for volatility spillovers in financial-sector assets. KO’s WEL ranges from $−0.0045\%$ to $−0.0021\%$, while CSCO’s lies between $−0.0022\%$ and $−0.0010\%$. DOW, as a reference index asset, shows the smallest utility gain, with WEL ranging from $−0.0020\%$ to $−0.0011\%$, given its lower idiosyncratic variance and high market beta.

\noindent These differences underscore the heterogeneous value of joint volatility modeling across sectors. Assets with higher tail risk, leverage effects, or volatility clustering show larger utility gaps relative to the constant-variance strategy. Importantly, the MHN–GARCH filter captures both systematic and idiosyncratic volatility shocks, producing forward-looking portfolio weights that respond intelligently to market dynamics. In contrast, the Merton benchmark applies static allocations regardless of market turbulence or opportunity, leading to underperformance and unnecessary utility loss.

\noindent The monotonic decrease in absolute WEL with increasing $\gamma$ also reveals that less risk-averse investors capture proportionally larger gains from adopting a dynamic strategy. These investors are more sensitive to expected return variation and less averse to temporary volatility, allowing them to capitalize more fully on the adaptive nature of the MHN–GARCH filter.

\noindent The wealth-equivalent loss analysis offers compelling empirical evidence for the economic relevance of dynamic, multivariate volatility filtering. By embedding filtered conditional variances and covariances into forward-looking optimization, the MHN–GARCH strategy materially improves investor welfare across a wide range of risk preferences and asset types. These results reinforce the central argument of the paper: that accounting for time-varying risk is not a secondary refinement but a fundamental requirement for optimal portfolio construction.

\subsection{Sensitivity of Optimal Allocation to Parameters} \label{sensitivityanalysis}

\noindent To assess the robustness of our model and its implications for portfolio decisions, we examine the sensitivity of optimal asset allocations to three key parameters in the MHN–GARCH framework: the correlation-driver parameter ($a$), the correlation coefficient ($\rho$), and the coefficient of relative risk aversion ($\gamma$). This analysis helps quantify how economically meaningful each parameter is and provides insight into how estimation errors or model misspecification might affect investor behavior.

\noindent Figures~\ref{fig:corr_driver}-\ref{fig:corr}-\ref{fig:gamma} display the sensitivity results for six representative assets in the DJIA: CVX, CSCO, KO, DOW, GS, and HD. For each asset, we track changes in the optimal allocation to both the stock and the index as the parameter in question varies across plausible ranges.

\subsubsection{Sensitivity to the Correlation-Driver Parameter ($a$)}

\begin{figure}[htbp]
    \centering
    \includegraphics[width=\textwidth,%
      height=\textheight,%
      keepaspectratio]{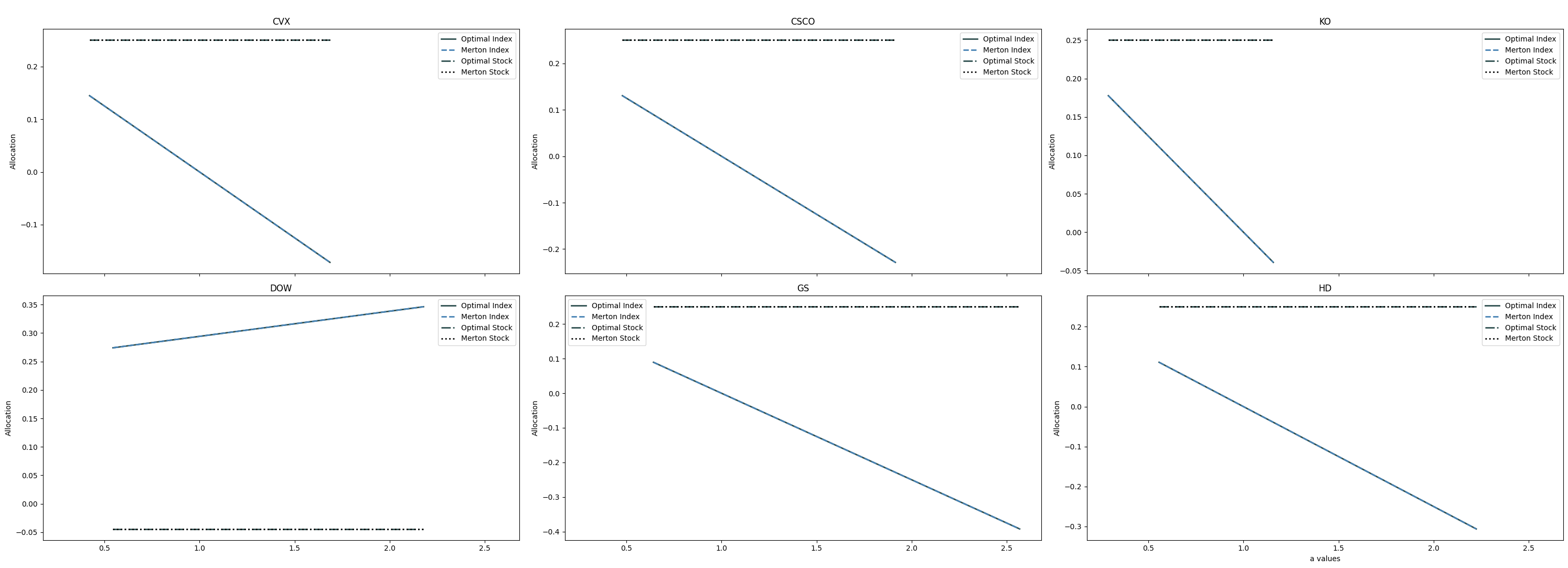}
    \caption{Sensitivity to Correlation Driver $a$}
    \label{fig:corr_driver}
\end{figure}

\noindent Figure~\ref{fig:corr_driver} illustrates how changes in the parameter a, which governs the transmission of volatility shocks from the index to individual assets, influence optimal portfolio weights. We observe strong, quasi-linear effects. For example, as a increases from 0.3 to 2.5, the index allocation for KO falls from 18\% to $–3\%$, while the stock allocation rises from 25\% to 72\%. CVX exhibits a similar pattern, with index weight declining from 14\% to $–17\%$ and stock weight increasing from 24\% to 75\%.

\noindent By contrast, DOW's allocations remain largely unchanged over the same range (index weight near 27\%, stock near $–4\%$), reflecting its central role in constructing the index and its limited idiosyncratic variance. These results highlight $a$’s importance in shaping optimal portfolio choices—larger a values amplify idiosyncratic volatility premia, shifting allocations toward the asset and away from the index. Investors highly exposed to correlation risk therefore benefit significantly from accurate estimation of this parameter.

\subsubsection{Senstivity to the Correlation Coefficient ($\rho$)}

\begin{figure}[htbp]
    \centering
    \includegraphics[width=\textwidth,%
      height=\textheight,%
      keepaspectratio]{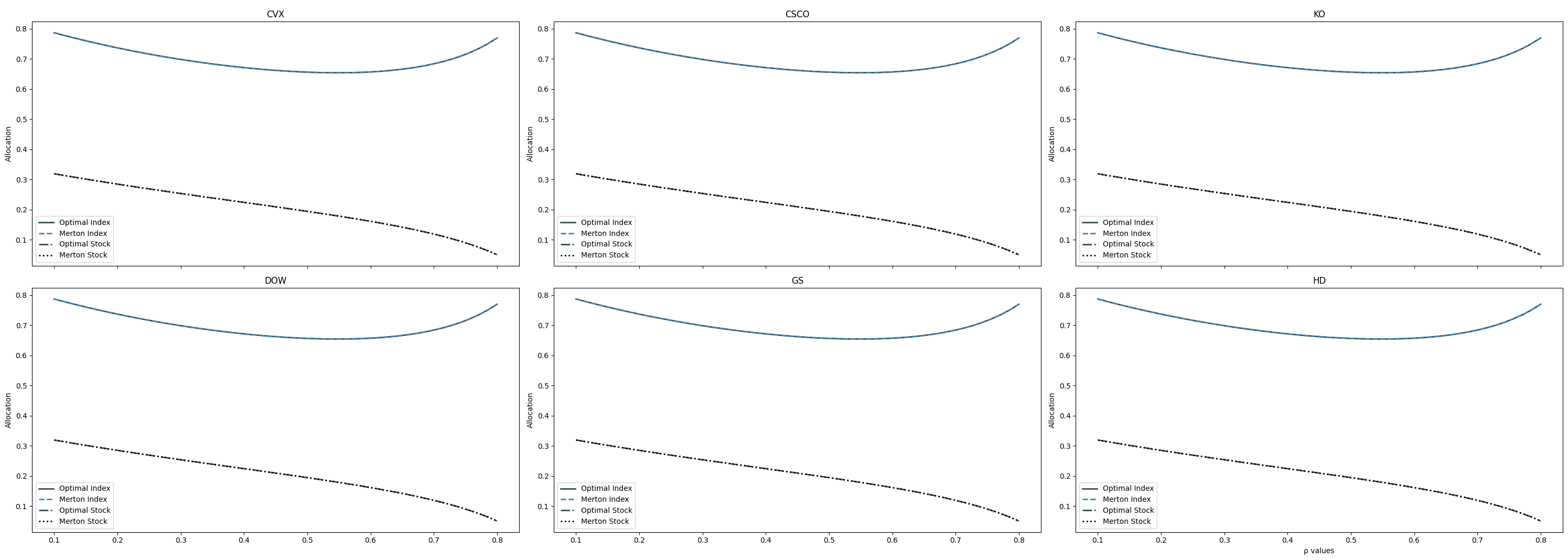}
    \caption{Sensitivity to Correlation $\rho$}
    \label{fig:corr}
\end{figure}

\noindent Figure~\ref{fig:corr} shows the effect of varying the unconditional correlation parameter $\rho$ from 0.1 to 0.8. In sharp contrast to the influence of a, these changes produce virtually no variation in optimal allocations. CVX’s index and stock weights remain fixed at approximately 14\% and 25\%, respectively; CSCO, KO, and HD show similarly flat profiles.

\noindent This insensitivity to $\rho$ reflects the model’s calibration: the baseline filtered correlation is already high (e.g., $\rho$ $\approx$ 0.83 for CVX and HD), so small perturbations around this level have negligible impact on the estimated conditional covariance matrix. The robustness of allocations to small shifts in $\rho$ is a desirable property, suggesting that the model remains stable under modest correlation misspecification.

\subsubsection{Sensitivity to the Coefficient of Constant Relative Risk Aversion ($\gamma$)}

\begin{figure}[htbp]
    \centering
    \includegraphics[width=\textwidth,%
      height=\textheight,%
      keepaspectratio]{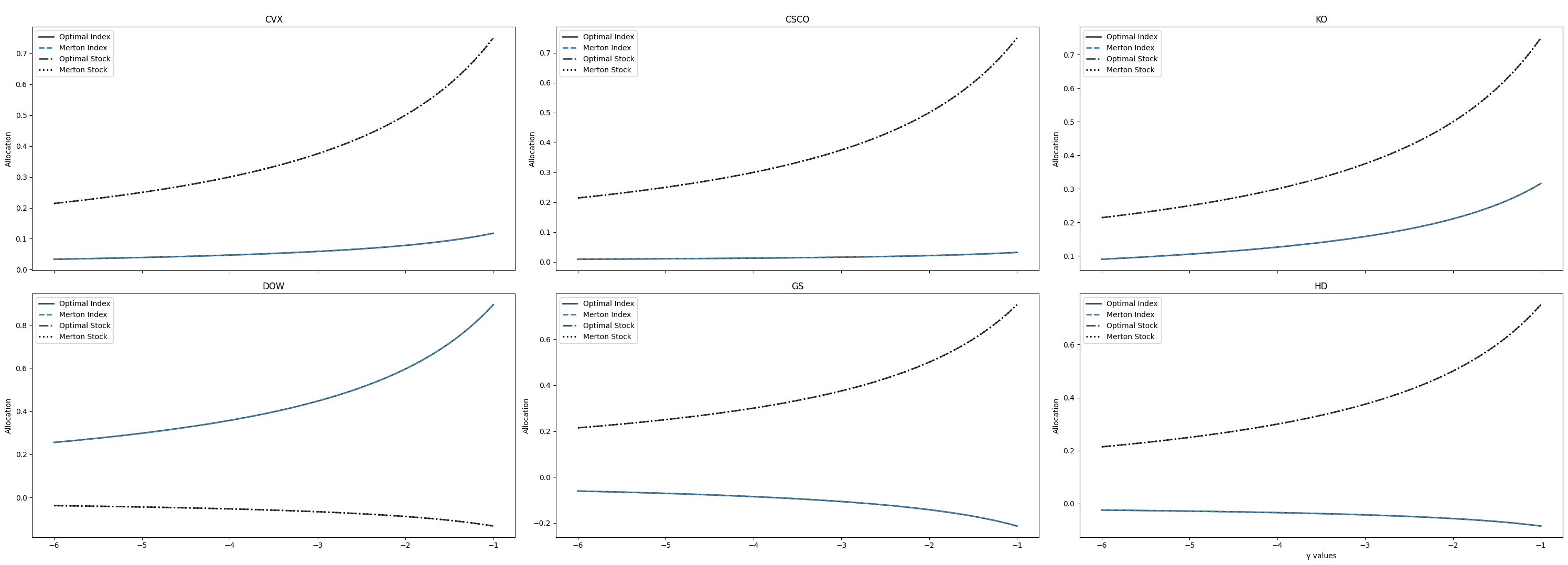}
    \caption{Sensitivity to Risk Aversion $\gamma$}
    \label{fig:gamma}
\end{figure}

\noindent Figure~\ref{fig:gamma} examines how changes in the investor’s risk aversion influence optimal portfolio weights. As $\gamma$ increases from 1 to 6, the index allocation monotonically declines while the stock allocation rises exponentially.

\noindent For instance, CVX’s index allocation falls from 22\% at $\gamma = 1$ to 3\% at $\gamma = 6$, while its stock weight climbs from 25\% to 45\%. GS displays an even more dramatic adjustment: index weight plunges from 22\% to $–4\%$, and stock allocation jumps from 21\% to 75\% over the same $\gamma$ range. These shifts reflect greater willingness among less risk-averse investors to take on idiosyncratic volatility in exchange for higher expected returns, especially when filtered innovations signal favorable tail behavior or leverage effects.

\noindent Interestingly, DOW (which represents the index itself) exhibits a narrower range of allocation responses, since the bivariate extension collapses to a quasi-univariate problem for this asset. As a result, index allocation varies from 26\% to 85\%, but the range is driven more by changes in portfolio concentration than cross-asset rebalancing.

\noindent Overall, this sensitivity analysis underscores that precise estimation of a and $\gamma$ is critical, particularly in environments with volatile market structure or diverse investor risk preferences. In contrast, errors in $\rho$ estimation have relatively minor effects on allocation. The economically meaningful variation in allocations with respect to $a$ and $\gamma$ validates both the flexibility and practical power of the MHN–GARCH framework—but also warns of the potential for model-driven misallocations if key parameters are poorly estimated.

\section{Conclusion} \label{conclusion}

\noindent This paper develops and empirically validates a multivariate affine GARCH(1,1) framework with Normal Inverse Gaussian (NIG) innovations to model joint volatility dynamics across asset returns. By pairing each of the 30 DJIA constituents with the DJIA index in a bivariate structure, we capture time-varying correlations, leverage effects, and heavy-tailed innovations that univariate or static models cannot accommodate. The resulting filtered variance–covariance matrix feeds directly into two key applications: dynamic portfolio optimization for a CRRA investor and closed-form option pricing based on an extended Heston–Nandi framework.

\noindent Our empirical findings support three principal conclusions. First, dynamic portfolio allocations derived from MHN–GARCH filtering deliver consistently higher expected utility relative to the Merton constant-variance benchmark, as reflected in strictly negative wealth-equivalent losses across all levels of risk aversion. Second, mean–variance efficient frontiers constructed using joint dynamic covariances dominate those based on univariate ARFIMA–FIGARCH filters, offering superior Sharpe ratios and risk–return trade-offs. Third, the extension of Heston–Nandi pricing to a multivariate setting produces empirically realistic implied volatility surfaces, replicating well-documented skew and smirk patterns in both call and put options.

\noindent Sensitivity analyses further underscore the model’s practical robustness. Optimal allocations respond meaningfully to variation in the correlation-driver parameter and investor risk aversion but remain stable under small changes to unconditional correlation. This calibration behavior highlights the model’s resilience under mild misspecification while emphasizing the importance of estimating key transmission mechanisms with care.

\noindent Beyond its empirical performance, the MHN–GARCH framework makes a methodological contribution by integrating multivariate volatility filtering, intertemporal portfolio choice, and derivative valuation under a unified, tractable structure. The model’s affine form permits closed-form expressions for both optimal asset weights and option prices, making it viable for large-scale implementation in institutional settings. Importantly, it addresses long-standing limitations in volatility modeling—specifically the inability to jointly handle tail risk, co-movement, and dynamic correlation in a scalable way.

\noindent Looking forward, several extensions present promising directions for future research. One avenue is to incorporate additional state variables—such as stochastic interest rates, macroeconomic indicators, or variance risk premia—into the affine filter. Another is to explore real-time updating using Bayesian or sequential learning methods to enhance resilience to structural breaks. Finally, applications to portfolio insurance, multi-asset derivative strategies, and stress testing across asset classes can further demonstrate the model’s flexibility in complex risk environments.

\noindent By unifying multivariate volatility filtering with forward-looking investment and pricing decisions, this framework offers a powerful tool set to manage joint asset risks under realistic market dynamics. The gains in investor utility, improved portfolio efficiency, and accurate option pricing documented in this study underscore the essential role of dynamic, non-Gaussian multivariate models in modern financial decision-making.

\newpage

\appendix

\centering \section*{\textbf{APPENDIX: Technical Implementation and Supporting Results}}

\RaggedRight

\section{Intertemporal Portfolio Choice} \label{ipc}

\noindent Let $W_t>0$ denote the investor’s wealth at time~$t$ and 
\[
\mathbf h_t
  \;=\;
  \bigl(h_{1,t},h_{2,t},\dots,h_{n,t}\bigr)^{\!\top}
\]
\noindent collect the $n$ conditional variances generated by the
MHN--GARCH$(1,1)$ recursion in equations described in section~\ref{MHN-GARCH Process}. The portfolio--weight vector is
\[
\boldsymbol\pi_t
  \;=\;
  \bigl(\pi_{0,t},\pi_{1,t},\dots,\pi_{n,t}\bigr)^{\!\top},
\qquad
\pi_{0,t}=1-\sum_{i=1}^{n}\pi_{i,t},
\]
\noindent where $\pi_{i,t}$ is the share of wealth invested in risky asset~$i$
and $\pi_{0,t}$ in the risk‑free asset. Let $\mathbf r_{t+1}=(r_{1,t+1},\dots,r_{n,t+1})^{\!\top}$ be the
one‑period log returns. With continuous compounding,
\begin{equation}
  W_{t+1}
  \;=\;
  W_t\,
  \exp\!\bigl(r+\boldsymbol\pi_t^{\!\top}\mathbf r_{t+1}\bigr),
\end{equation}

\noindent where $r$ is the risk‑free rate. The investor has CRRA utility with coefficient $\gamma>0$,
\(
U(W)=\tfrac{W^{1-\gamma}}{1-\gamma}.
\)
The value function is defined as:

\begin{equation}
V_t(W_t,\mathbf h_t)
  \;=\;
  \max_{\{\boldsymbol\pi_\tau\}_{\tau=t}^{T-1}}
  \mathbb E_t\!\bigl[U(W_T)\bigr].
\end{equation}

\noindent By homotheticity, conjecture
\begin{equation}
  V_t(W_t,\mathbf h_t)
  \;=\;
  \frac{W_t^{1-\gamma}}{1-\gamma}\,
  \exp\!\Bigl(
      D_t+\sum_{i=1}^{n}E_{i-1,t}\,h_{i,t}
      \Bigr),
  \label{eq:V-form}
\end{equation}

\noindent with deterministic functions
$D_t,E_{0,t},\dots,E_{n-1,t}$. 
\noindent Substituting (\ref{eq:V-form}) into the Bellman equation yields
\begin{align}
&\max_{\boldsymbol\pi_t}
  \log
  \mathbb E_t\!\Bigl[
    \exp\bigl(
      (1-\gamma)\bigl[r+\boldsymbol\pi_t^{\!\top}\mathbf r_{t+1}\bigr]
      +D_{t+1}
      +\sum_{i=1}^{n}E_{i-1,t+1}\,h_{i,t+1}
    \bigr)
  \Bigr].
\end{align}

\noindent Conditional log‑normality implies the first‑order conditions

\begin{align}
\pi_{i,t}^{\star}
  &=
  \frac{\bigl(\tfrac12+\lambda_i h_{i,t}\bigr)
        \bigl[1-2\alpha_iE_{i,t+1}\bigr]
        -2\alpha_i\theta_iE_{i,t+1}}
       {1-2\alpha_iE_{i,t+1}-\gamma},
  \qquad i=2,\dots,n,\\[4pt]
\pi_{1,t}^{\star}
  &=
  \frac{\bigl(\tfrac12+\lambda_1 h_{1,t}\bigr)
        \bigl[1-2\alpha_1E_{0,t+1}\bigr]
        -2\alpha_1\theta_1E_{0,t+1}}
       {1-2\alpha_1E_{0,t+1}-\gamma}
  -\sum_{i=2}^{n}a_i\,\pi_{i,t}^{\star}.
\end{align}

\noindent Matching coefficients of $h_{i,t}$ and the constant in the Bellman equation gives

\begin{align}
E_{0,t}
 &= (\beta_1+\alpha_1\theta_1^{2})E_{0,t+1}
    +G_0\!\bigl(E_{t+1},\boldsymbol\pi_t^{\star}\bigr),\\[2pt]
E_{i,t}
 &= (\beta_i+\alpha_i\theta_i^{2})E_{i,t+1}
    +G_i\!\bigl(E_{t+1},\boldsymbol\pi_t^{\star}\bigr),
    \quad i=1,\dots,n-1,\\[2pt]
D_t
 &= D_{t+1}
    +\gamma r
    +\omega_1E_{0,t+1}
    +\sum_{i=1}^{n-1}\omega_{i+1}E_{i,t+1}
    -\tfrac12\sum_{i=0}^{n-1}
       \log\!\bigl[1-2\alpha_{i+1}E_{i,t+1}\bigr],
\end{align}

\noindent with terminal conditions $E_{i,T}=0$ and $D_T=0$. The auxiliary functions $G_i(\cdot)$ correspond to the closed‑form expressions implemented in backward recursion, proved in \citet{escobar2025}.

\noindent Given optimal weights $\boldsymbol\pi_t^{\star}$, wealth evolves as
\begin{equation}
  W_{t+1}
  \;=\;
  W_t\,
  \exp\!\bigl(r+\boldsymbol\pi_t^{\star\top}\mathbf r_{t+1}\bigr),
  \qquad
  W_0=1,
\end{equation}

\noindent producing the \emph{forward‐looking} path
$\{W_t\}_{t=0}^{T}$. In contrast to the time-varying allocation approach, we also compute the Merton constant variance benchmark to demonstrate the tractability of the time-varying optimal solution with the theory of intertemporal risk sharing in portfolio optimization. Assume $\mathbf r_{t+1}$ is i.i.d.\ with mean $\boldsymbol\mu$ and covariance $\Sigma$. The investor chooses a \emph{constant} weight vector $\boldsymbol\pi$ to maximize
$\mathbb E\bigl[U(W_T)\bigr].$ Since $
\log W_T
  = \log W_0
  +\sum_{t=0}^{T-1}
     \bigl[r+\boldsymbol\pi^{\!\top}\mathbf r_{t+1}\bigr],
$ the optimization can be simplified to
\begin{equation}
  \max_{\boldsymbol\pi}\;
  (1-\gamma)
  \bigl[
    r+\boldsymbol\pi^{\!\top}\boldsymbol\mu
    -\tfrac12\gamma\,\boldsymbol\pi^{\!\top}\Sigma\boldsymbol\pi
  \bigr].
\end{equation}

\noindent First‑order necessary conditions give
\begin{equation}
      \boldsymbol\pi^{\star}_{\text{M}}
      \;=\;
      \frac{1}{\gamma}\,\Sigma^{-1}\boldsymbol\mu
\end{equation}

Wealth then follows
\[
  W_{t+1}^{\text{M}}
  = W_t^{\text{M}}
    \exp\!\bigl(r+\boldsymbol\pi_{\text{M}}^{\!\top}\mathbf r_{t+1}\bigr),
  \qquad
  W_0^{\text{M}}=1.
\]

\bigskip
\noindent The MHN–GARCH optimal policy rule adapts in dynamically to shocks in
$\mathbf h_t$, capturing both idiosyncratic and
systemic volatility, whereas the Merton strategy is static and
therefore ignores the intertemporal hedging opportunities
created by time‑varying joint risk.

\section{Standardized Innovations} \label{innov}

\noindent In equations (1) to (3) in section~\ref{MHN-GARCH Process}, the standardized NIG innovations are given by:

\begin{align}
z_{n,t} &= \frac{X_{n,t} - \left[X_{n,t-1} + r + \lambda_{1,n}\, h_{1,t} + \lambda_n\, h_{n,t} + a_{n}\, \sqrt{h_{1,t}}\, z_{1,t}\right]}{\sqrt{h_{n,t}}},\;\;n=1,2,\dots,
\end{align}

\noindent where $X_{1,t}$ is the log-price process of the index and $X_{n,t}$ is the log-price process of asset $n$. Additionally, $r$ is the risk-free rate, $\lambda_{1}$ is the market-price of the index, $\lambda_{n}$ is the market-price of asset $n$, and $\lambda_{1,2}$ is the co-market price of risk identifying the joint risk of the bivariate trading pair. $a_{n}$ is the correlation driver parameter uniquely identifying the correlation of the asset $n$'s volatility with that of the index. $h_{1,t}$ is the conditional volatility of the index whereas $h_{n,t}$ is the conditional volatility of asset $n$.

\begin{figure}[htbp]
    \centering
    \includegraphics[width=\textwidth,%
      height=\textheight,%
      keepaspectratio]{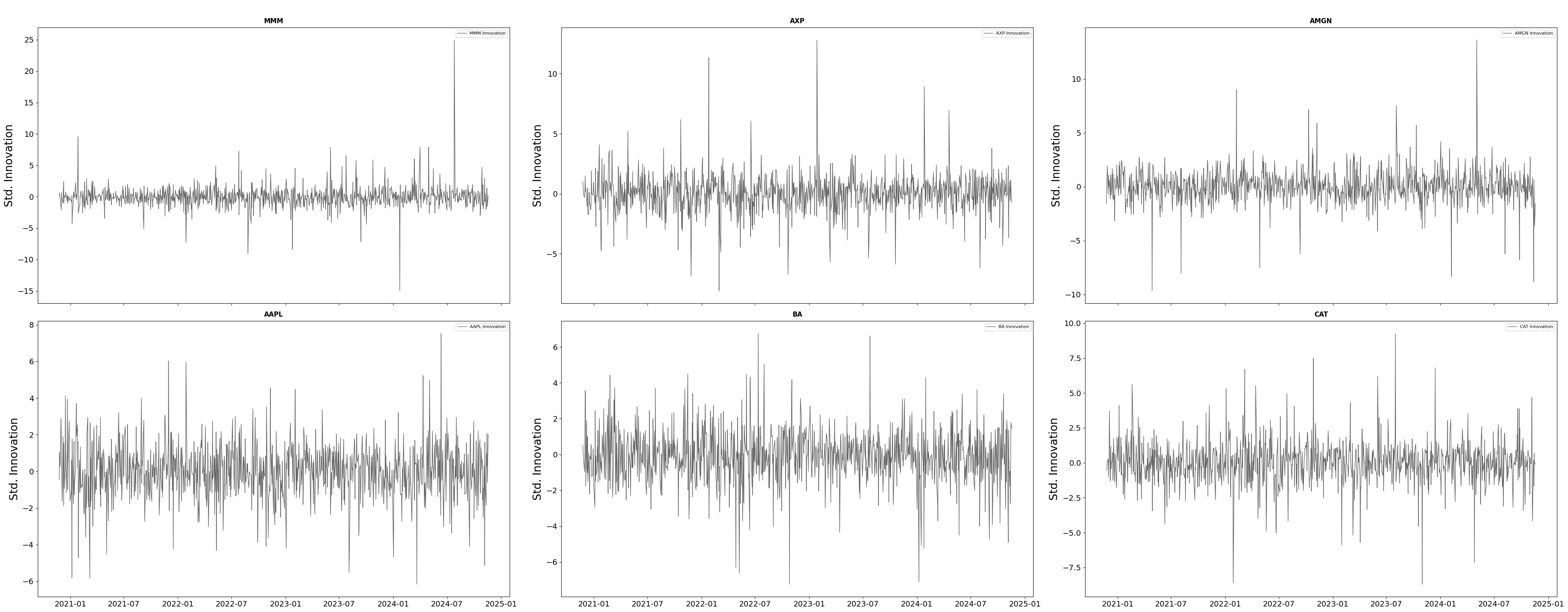}
    \label{fig:innovations1}
\end{figure}

\begin{figure}[htbp]
    \centering
    \includegraphics[width=\textwidth,%
      height=\textheight,%
      keepaspectratio]{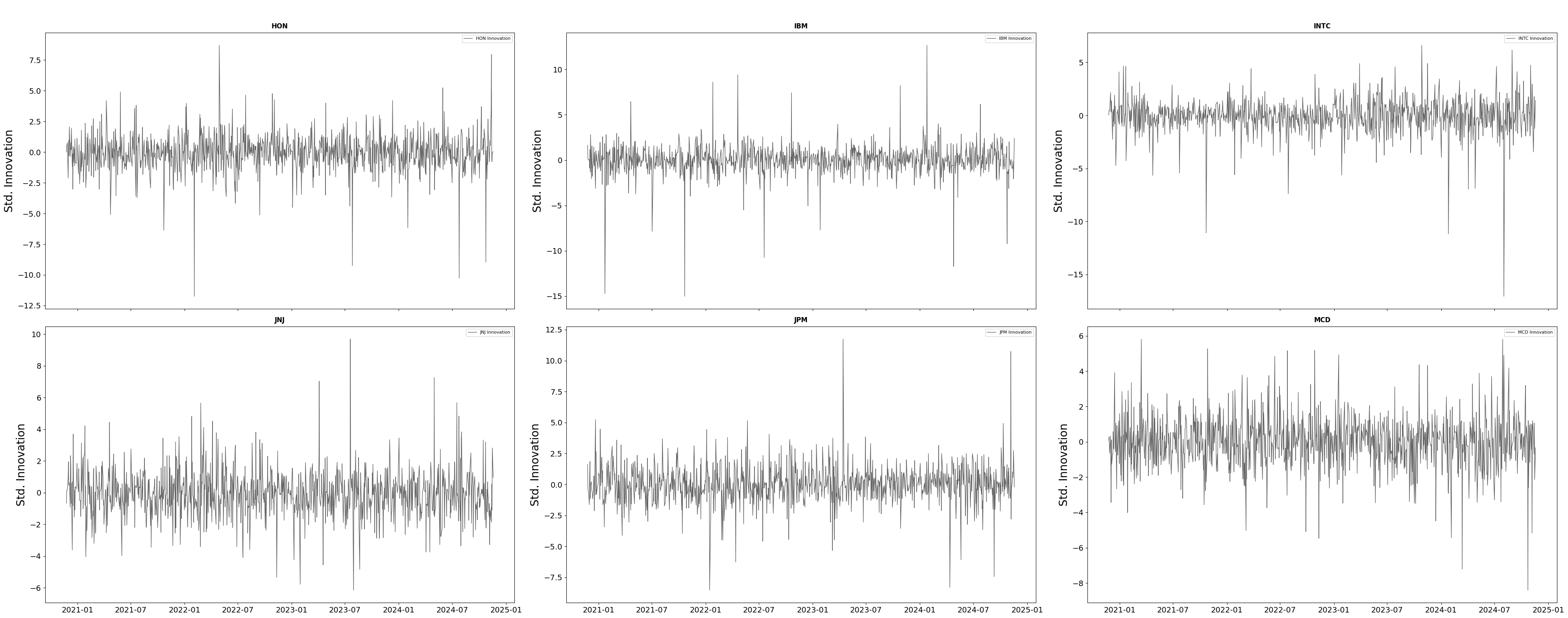}
    \label{fig:innovations3}
\end{figure}

\begin{figure}[htbp]
    \centering
    \includegraphics[width=\textwidth,%
      height=\textheight,%
      keepaspectratio]{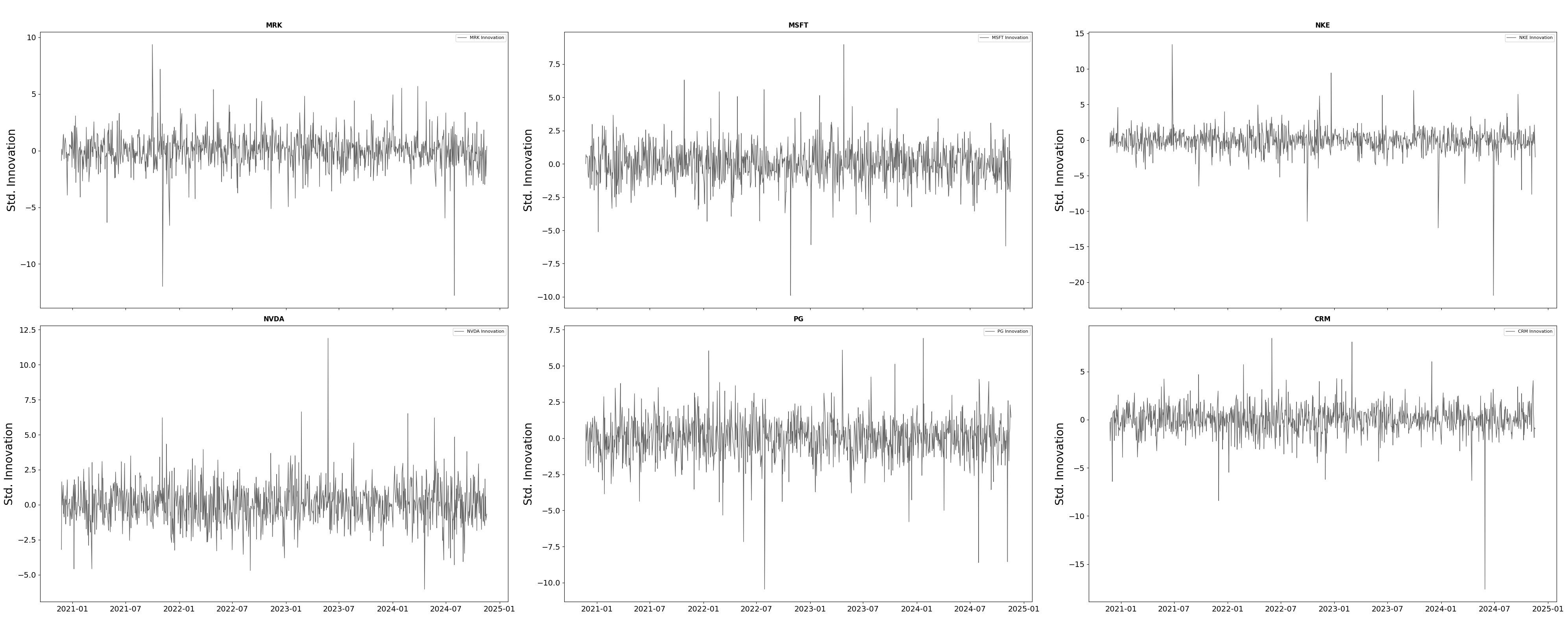}
    \label{fig:innovations4}
\end{figure}

\begin{figure}[htbp]
    \centering
    \includegraphics[width=\textwidth,%
      height=\textheight,%
      keepaspectratio]{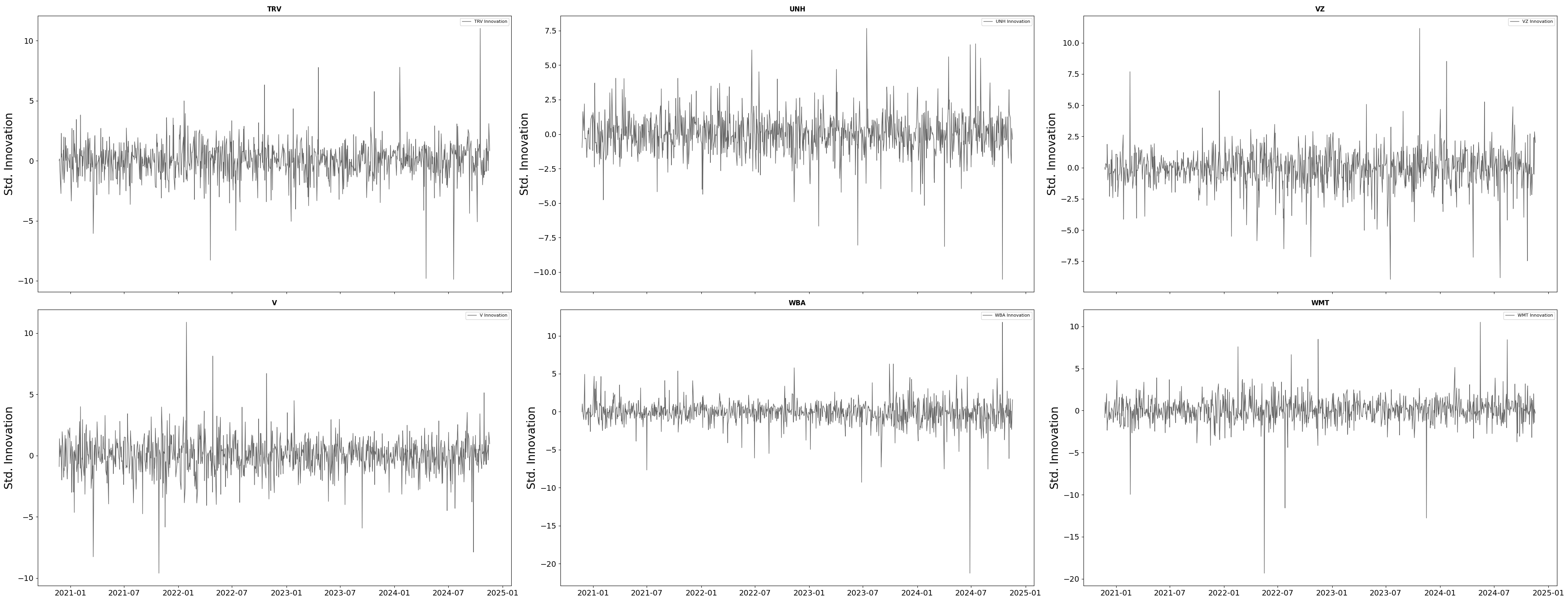}
    \caption{NIG innovations (asset‑specific unobservables) of Other Assets in the Portfolio}
    \label{fig:innovations5}
\end{figure}

\newpage

\section{Model-Based Volatility} \label{vol}

\noindent We plot the conditional volatility from the MHN-GARCH (1,1) specification, univariate ARFIMA(1,$d(m)$,1)-FIGARCH(1,$d(v)$,1), and historical volatility, as specified in section~\ref{MHN-GARCH Process}, for other assets in the portfolio to demonstrate the applicability of the model to a large universe of assets. Here, Dow-30 representative investor can filter volatilities for each asset without relying on any univariate structure and observe the volatility from the bivariate trading pair of the index and asset $n$. This demonstrates the general applicability of the algorithm and comparison of different estimates of volatilities across any number of assets in a portfolio of an investor.

\begin{figure}[htbp]
    \centering
    \includegraphics[width=\textwidth,%
      height=\textheight,%
      keepaspectratio]{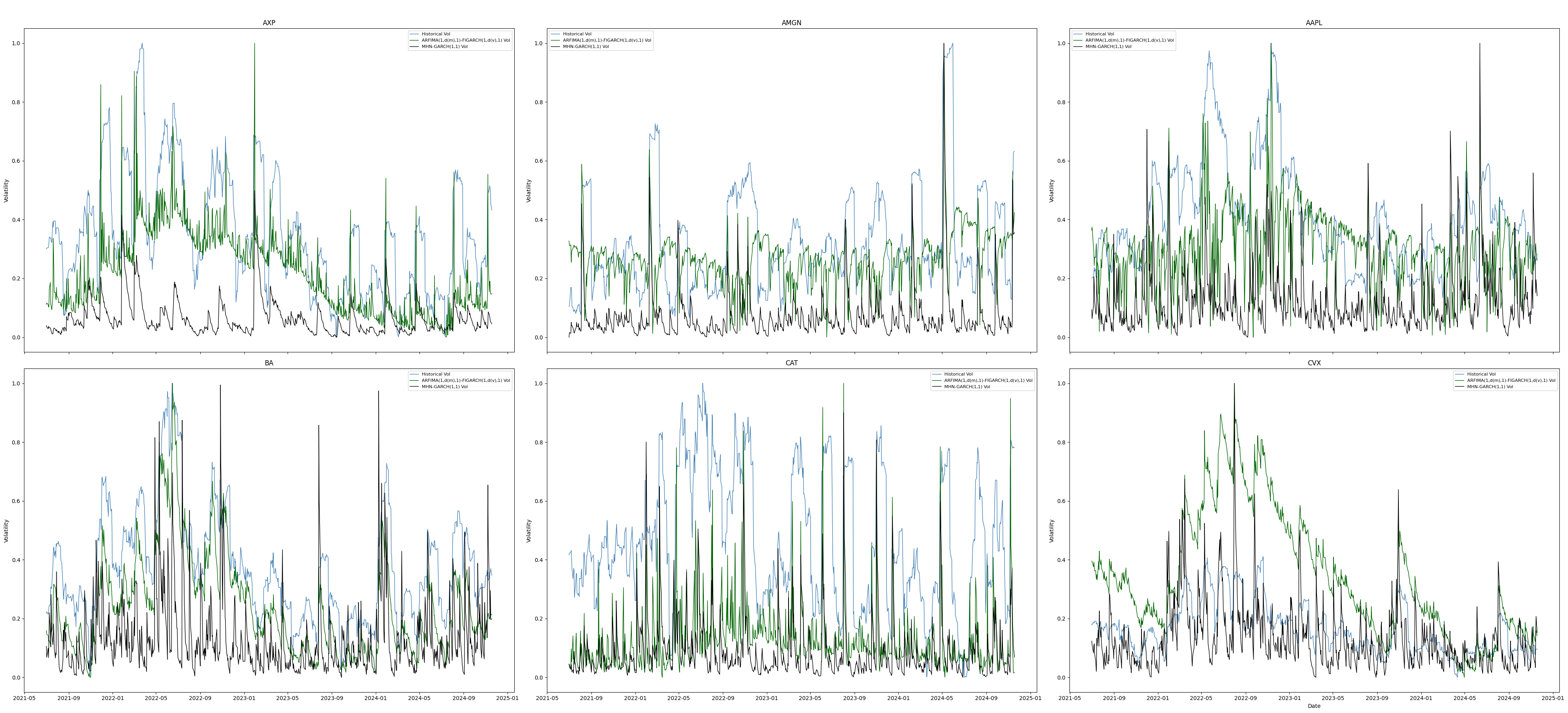}
    \label{fig:mbv1}
\end{figure}

\begin{figure}[htbp]
    \centering
    \includegraphics[width=\textwidth,%
      height=\textheight,%
      keepaspectratio]{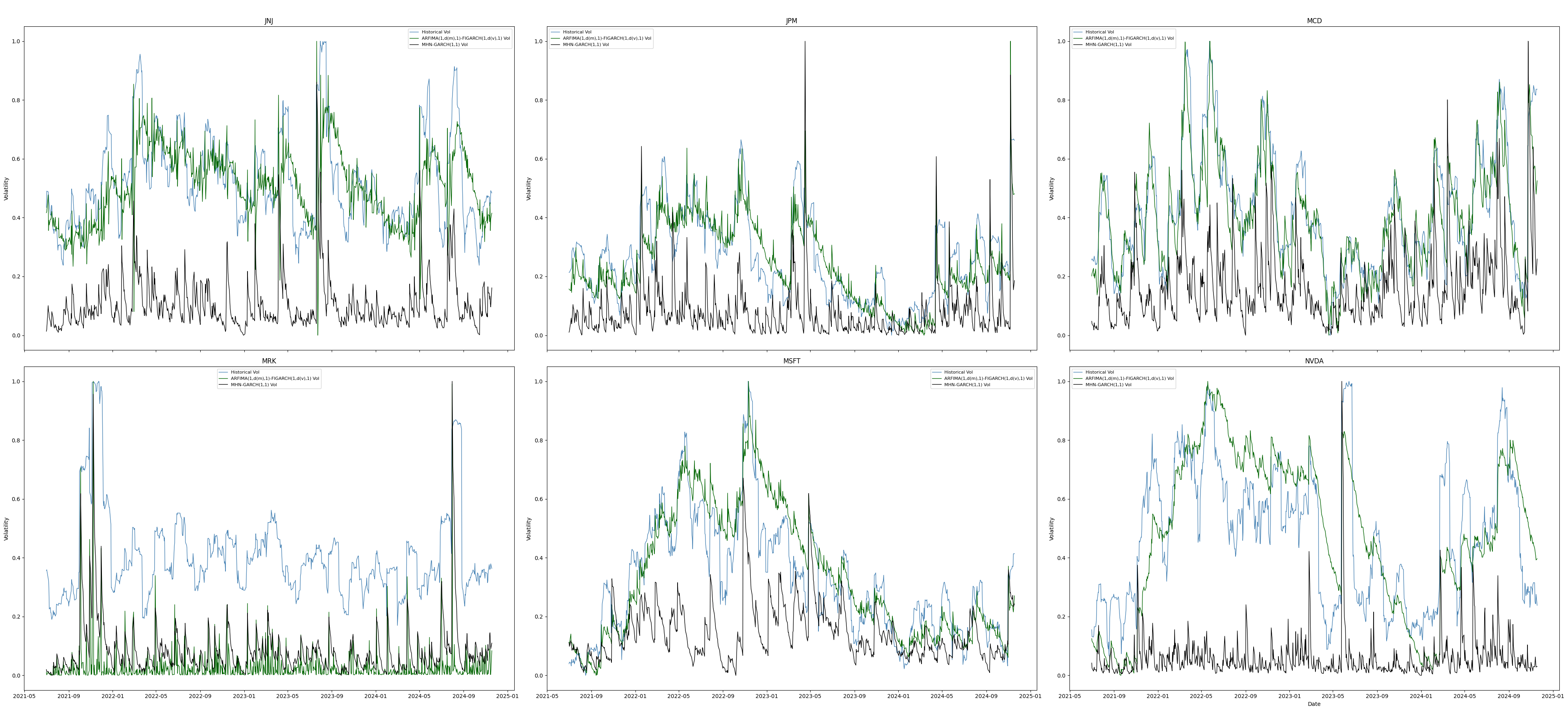}
    \label{fig:mbv3}
    \caption{Model-Based Volatility for Other Assets in the Portfolio}
\end{figure}

\newpage

\section{Call and Put Price Surfaces} \label{op}

\noindent We plot the call and put price surfaces from the option pricing specification detailed in section~\ref{options} for other assets in the portfolio. The smile and smirk behavior of the assets' call and put prices are demonstrated to highlight the general applicability of the MHN-GARCH (1,1) algorithm and the ability of using conditional volatility and parameter calibrations into the option pricing framework as described by \citet{hn2000}. 

\begin{figure}[htbp]
    \centering
    \includegraphics[width=\textwidth,%
      height=\textheight,%
      keepaspectratio]{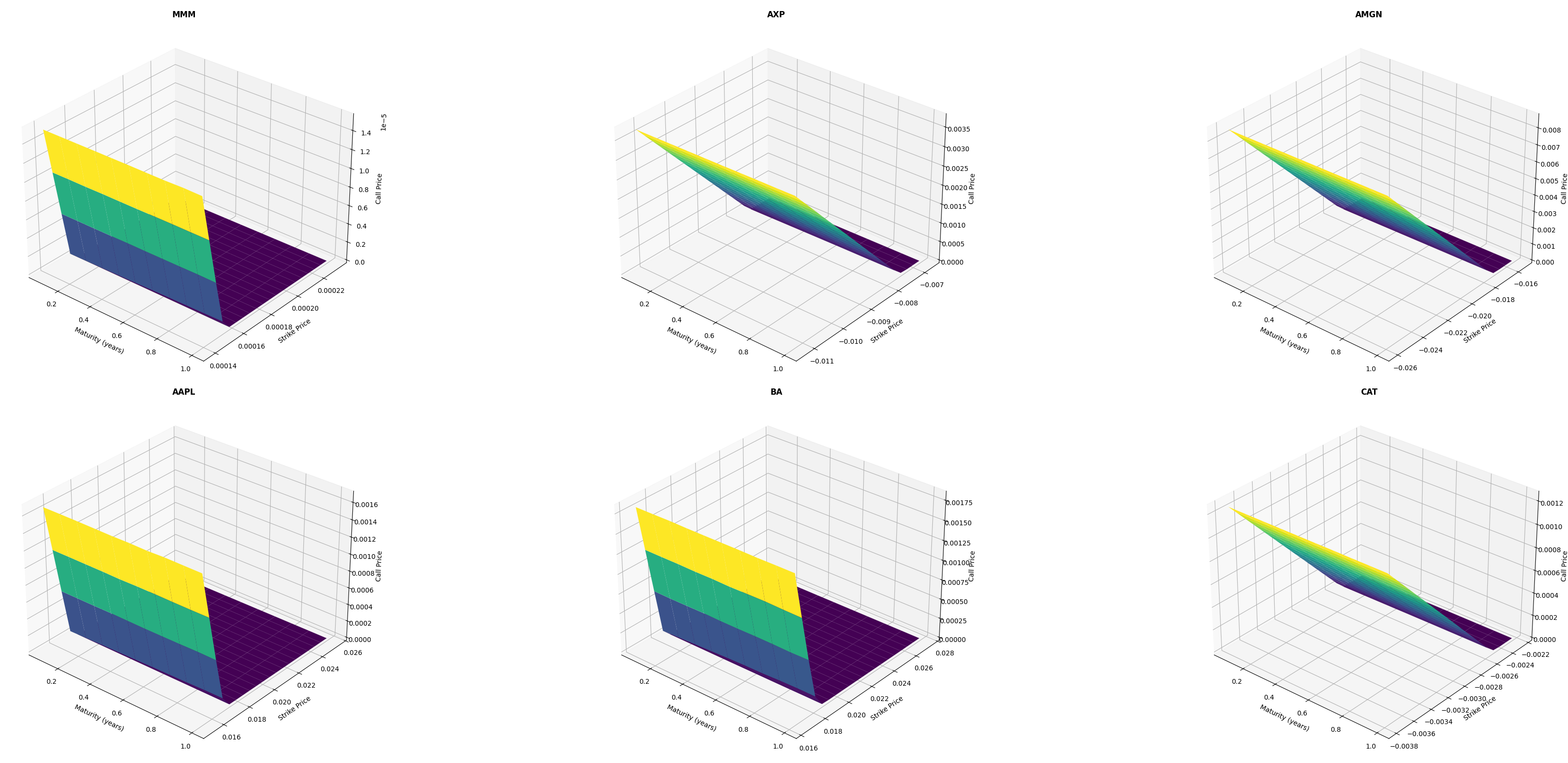}
    \caption{Call Price}
\end{figure}
\begin{figure}[htbp]
    \centering
    \includegraphics[width=\textwidth,%
      height=\textheight,%
      keepaspectratio]{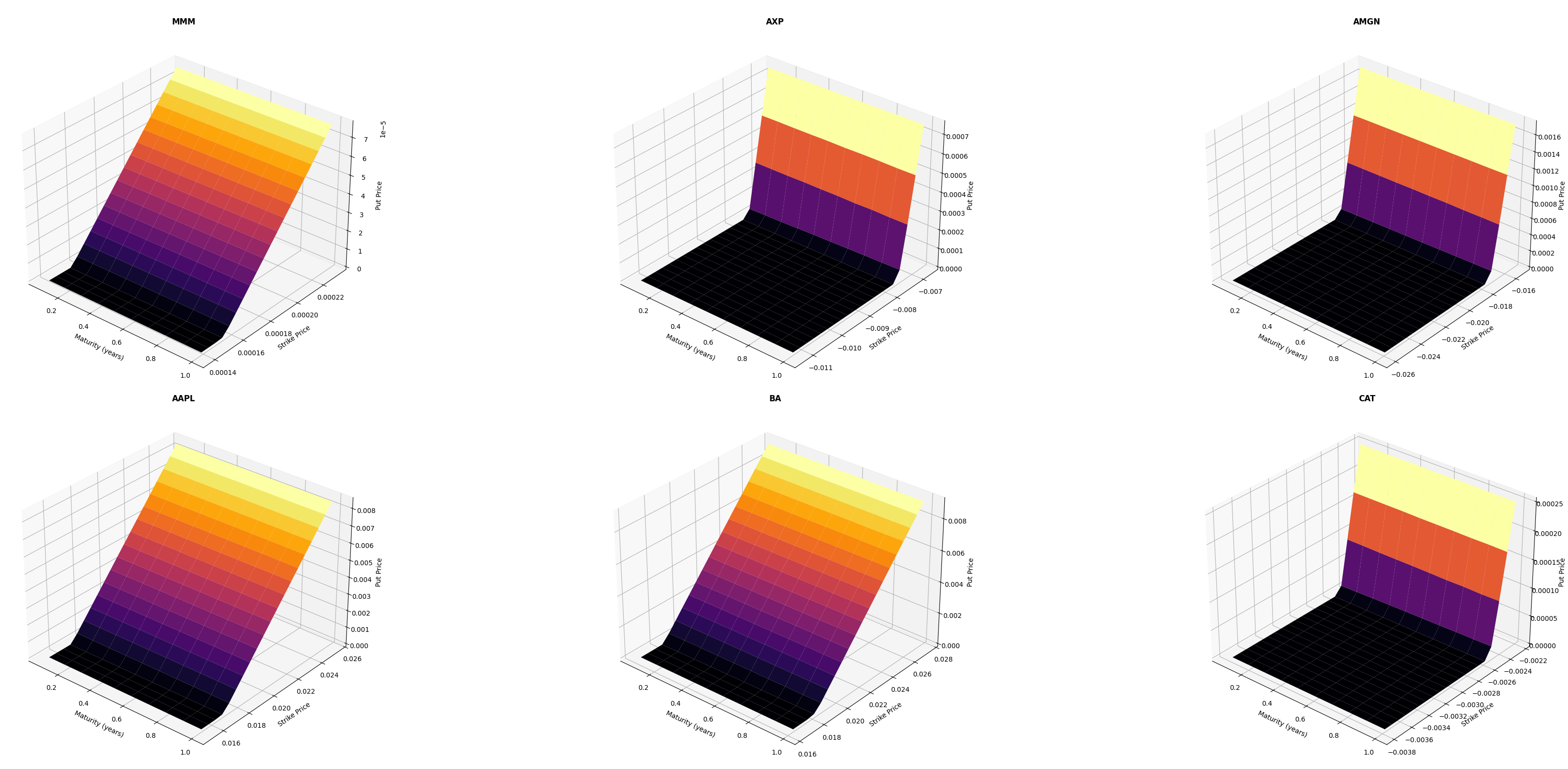}
      \caption{Put Price}
\end{figure}
\begin{figure}[htbp]
    \centering
    \includegraphics[width=\textwidth,%
      height=\textheight,%
      keepaspectratio]{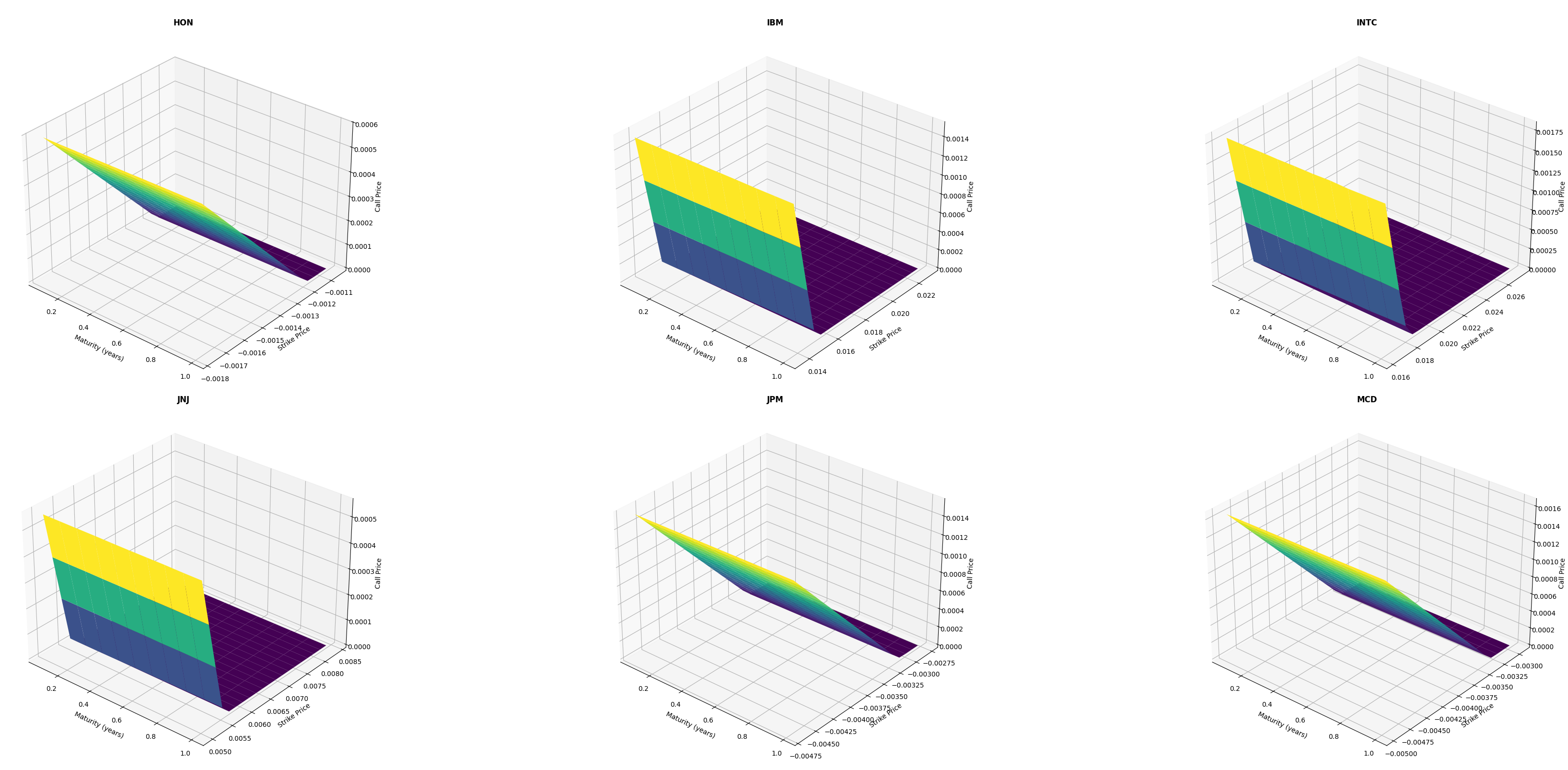}
      \caption{Call Price}
\end{figure}
\begin{figure}[htbp]
    \centering
    \includegraphics[width=\textwidth,%
      height=\textheight,%
      keepaspectratio]{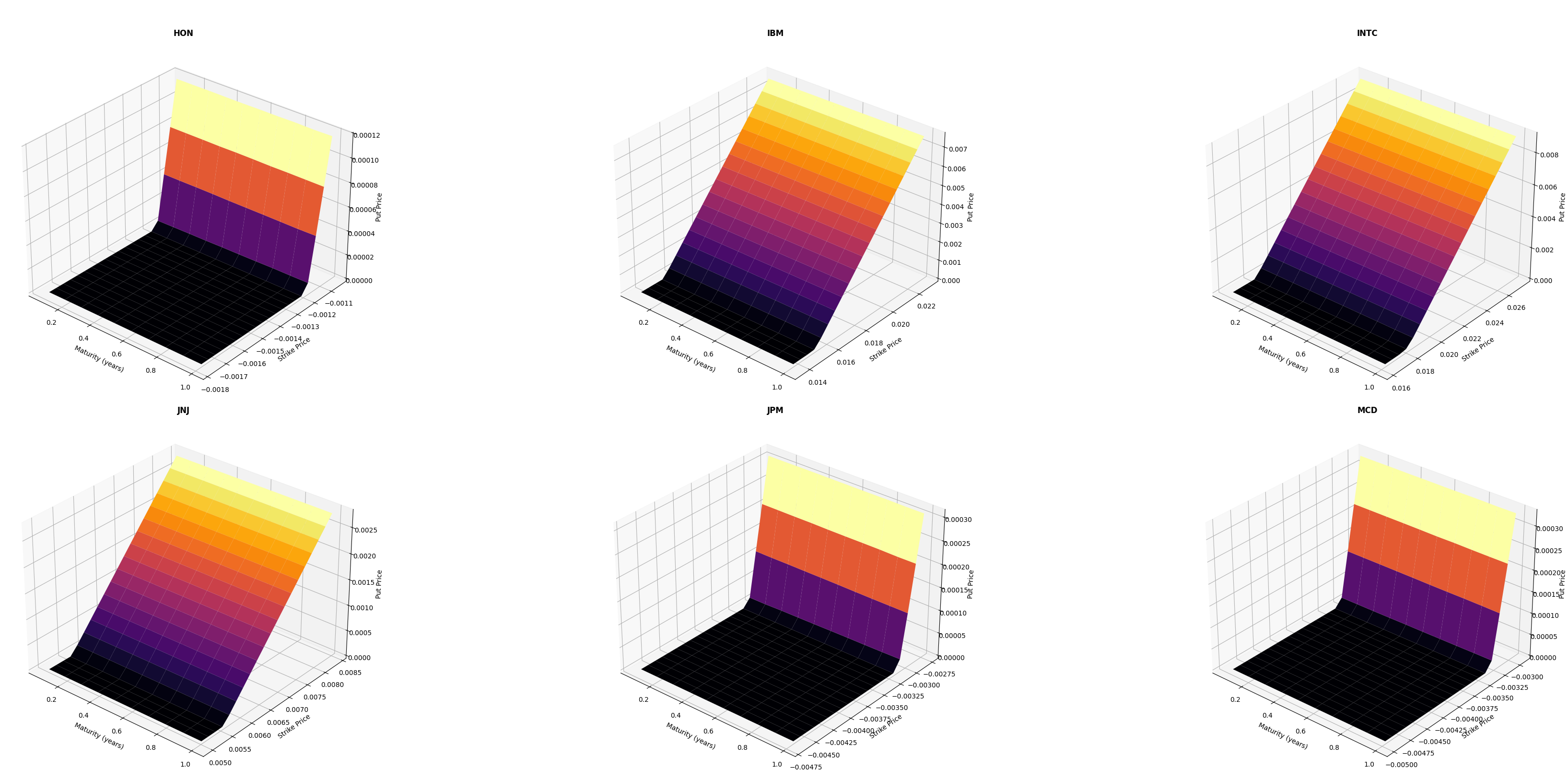}
      \caption{Put Price}
\end{figure}
\begin{figure}[htbp]
    \centering
    \includegraphics[width=\textwidth,%
      height=\textheight,%
      keepaspectratio]{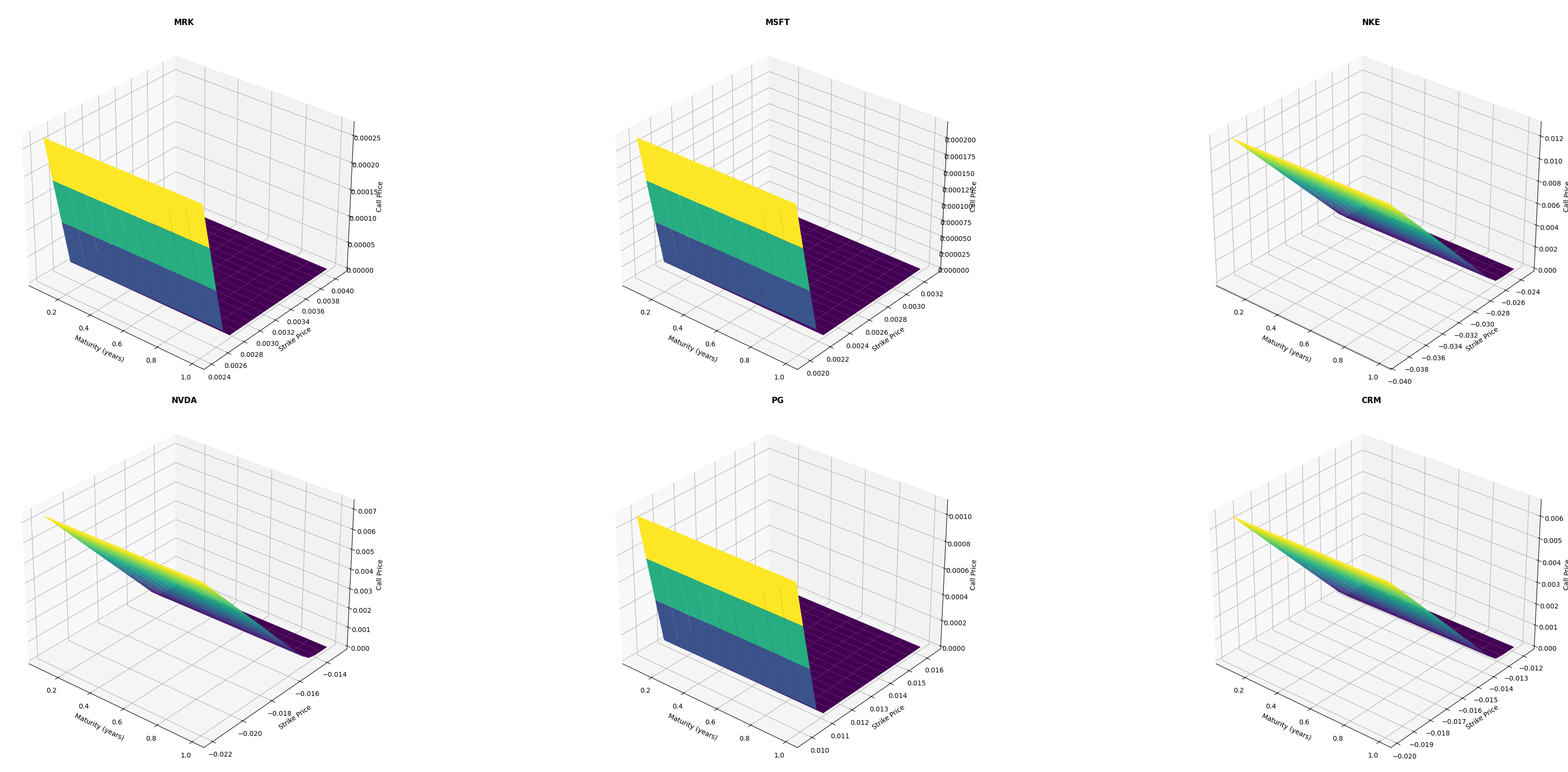}
      \caption{Call Price}
\end{figure}
\begin{figure}[htbp]
    \centering
    \includegraphics[width=\textwidth,%
      height=\textheight,%
      keepaspectratio]{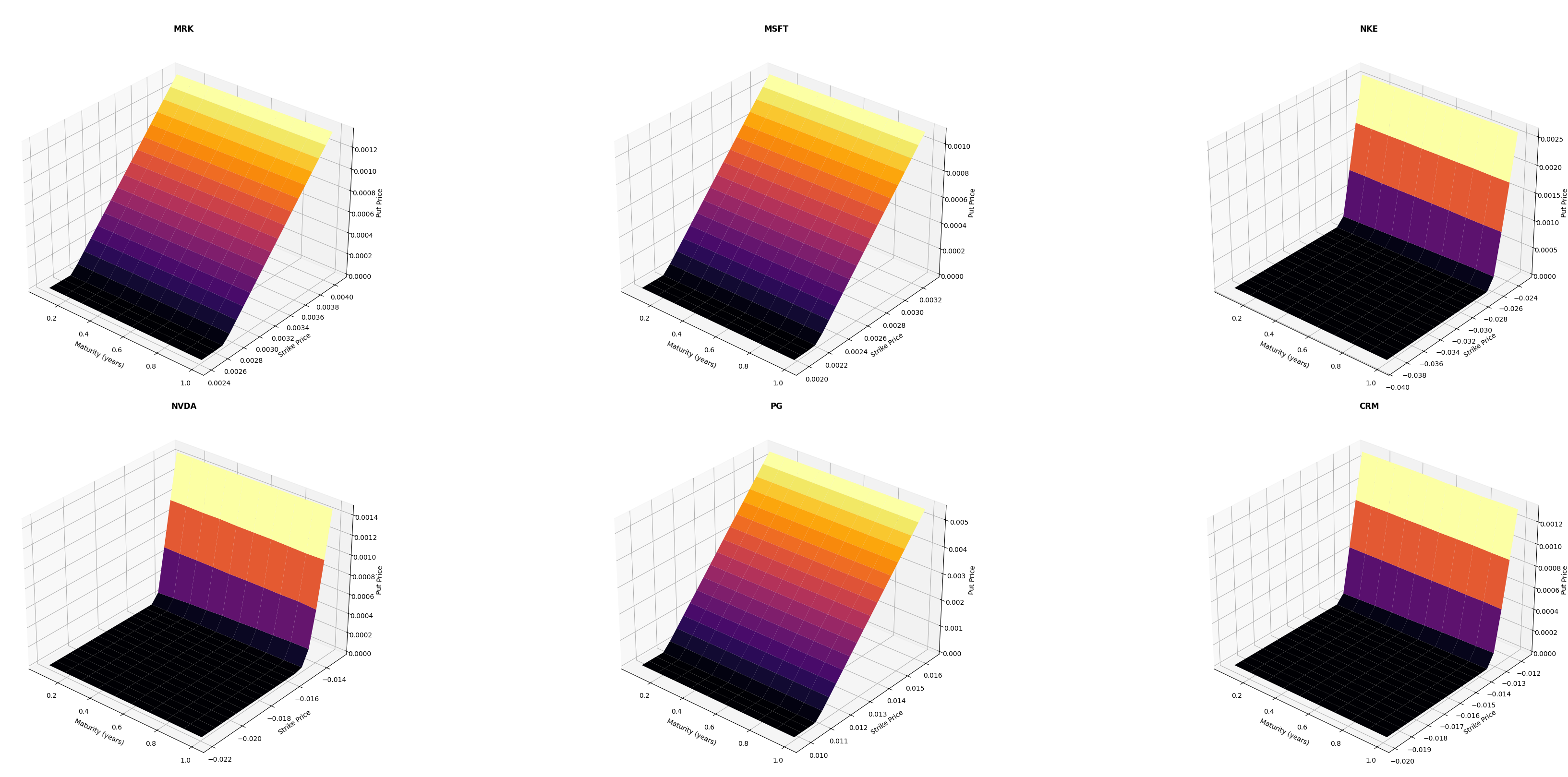}
      \caption{Put Price}
\end{figure}
\begin{figure}[htbp]
    \centering
    \includegraphics[width=\textwidth,%
      height=\textheight,%
      keepaspectratio]{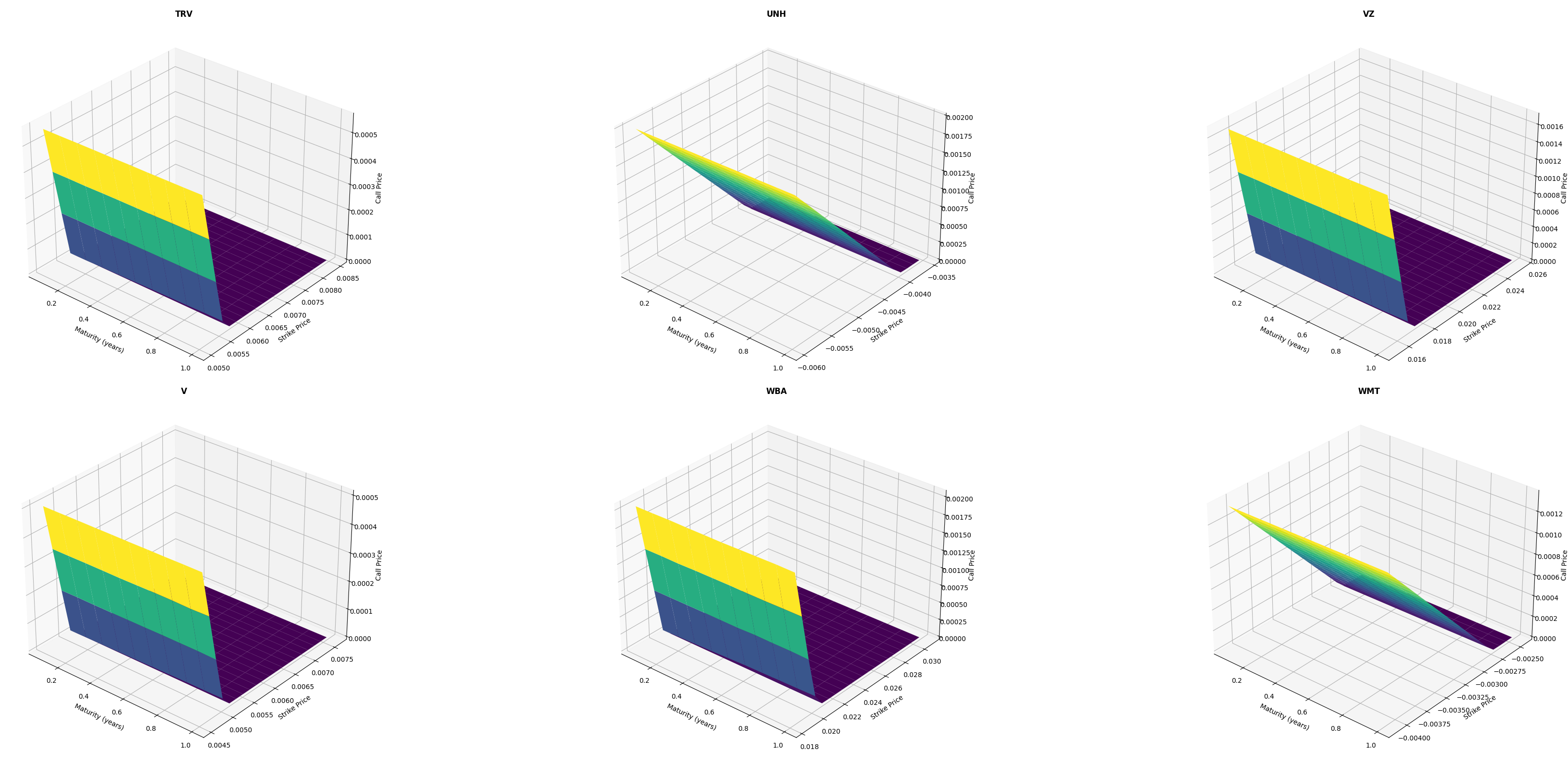}
      \caption{Call Price}
\end{figure}
\begin{figure}[htbp]
    \centering
    \includegraphics[width=\textwidth,%
      height=\textheight,%
      keepaspectratio]{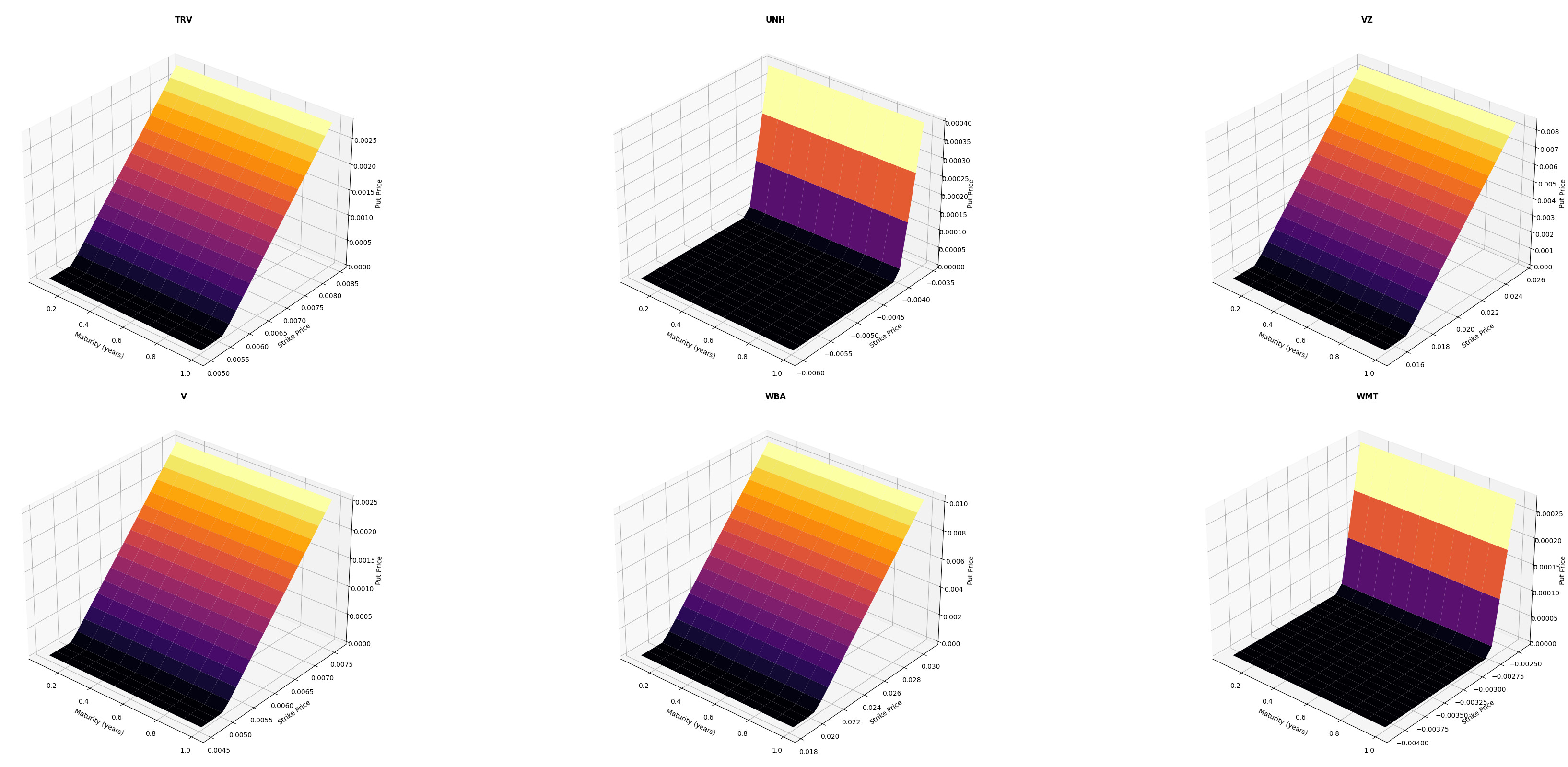}
      \caption{Put Price}
\end{figure}

\newpage

\section{CVaR Efficient Frontier} \label{CVaR}

\noindent We follow \citet{RockafellarUryasev2000} and \citet{McNeilFreyEmbrechts2015} for Conditional Value-at-Risk portfolio optimization. Consider a set of \(S\) return scenarios for \(N_k\) assets at iteration \(k\), captured in the matrix \(\mathbf{R}_k \in \mathbb{R}^{S \times N_k}\). Each row \(\mathbf{R}_k[s,:]\) corresponds to the returns of all assets under scenario \(s\). A portfolio weight vector \(\mathbf{w}_k \in \mathbb{R}^{N_k}\) must satisfy the constraints 

\[
\mathbf{w}_k \geq \mathbf{0}, 
\quad 
\mathbf{w}_k^\top \mathbf{1} = 1,
\]

reflecting non-negativity and full investment. Under scenario \(s\), the portfolio’s realized return (or loss if negative) is given by 

\[
r_{s,k} \;=\; \mathbf{R}_k[s,:] \,\mathbf{w}_k.
\]

To incorporate CVaR at the \(\alpha\) level (e.g., \(\alpha = 0.99\)), the optimization problem can be formulated using an auxiliary variable \(t\) and slack variables \(\{z_s\}_{s=1}^S\). In one common linear-programming representation, minimizing CVaR translates into minimizing

\[
t 
\;+\;
\frac{1}{(1-\alpha)\,S} \sum_{s=1}^S z_s,
\]
subject to
\[
z_s \;\ge\; -\,r_{s,k} \;-\; t, 
\quad
z_s \;\ge\; 0 
\quad\text{for all } s=1, \ldots, S,
\]

where \(-\,r_{s,k}\) represents a loss if \(r_{s,k}\) is negative. The decision variables \(\mathbf{w}_k\), \(t\), and \(\{z_s\}\) are all solved simultaneously, yielding the optimal minimum-CVaR portfolio \(\mathbf{w}_k^{\star}\). Because \(\mathbf{w}_k\) must also satisfy non-negativity and the budget constraint, the final optimization system becomes

\[
\begin{aligned}
\min_{\substack{\mathbf{w}_k,\,t,\,\{z_s\}}} 
&\;
t 
\;+\;
\frac{1}{(1-\alpha)\,S} \sum_{s=1}^S z_s,\\
\text{subject to}\;\;
& z_s \;\ge\; -\,r_{s,k} \;-\; t, \;\; z_s \;\ge\; 0, \quad s=1,\ldots,S,\\
& \mathbf{w}_k \;\ge\; \mathbf{0}, 
\quad
\mathbf{w}_k^\top \mathbf{1} \;=\; 1.
\end{aligned}
\]

\begin{figure}[htbp]
    \centering
    \begin{minipage}[t]{0.49\textwidth} 
        \centering
        \includegraphics[width=\textwidth]{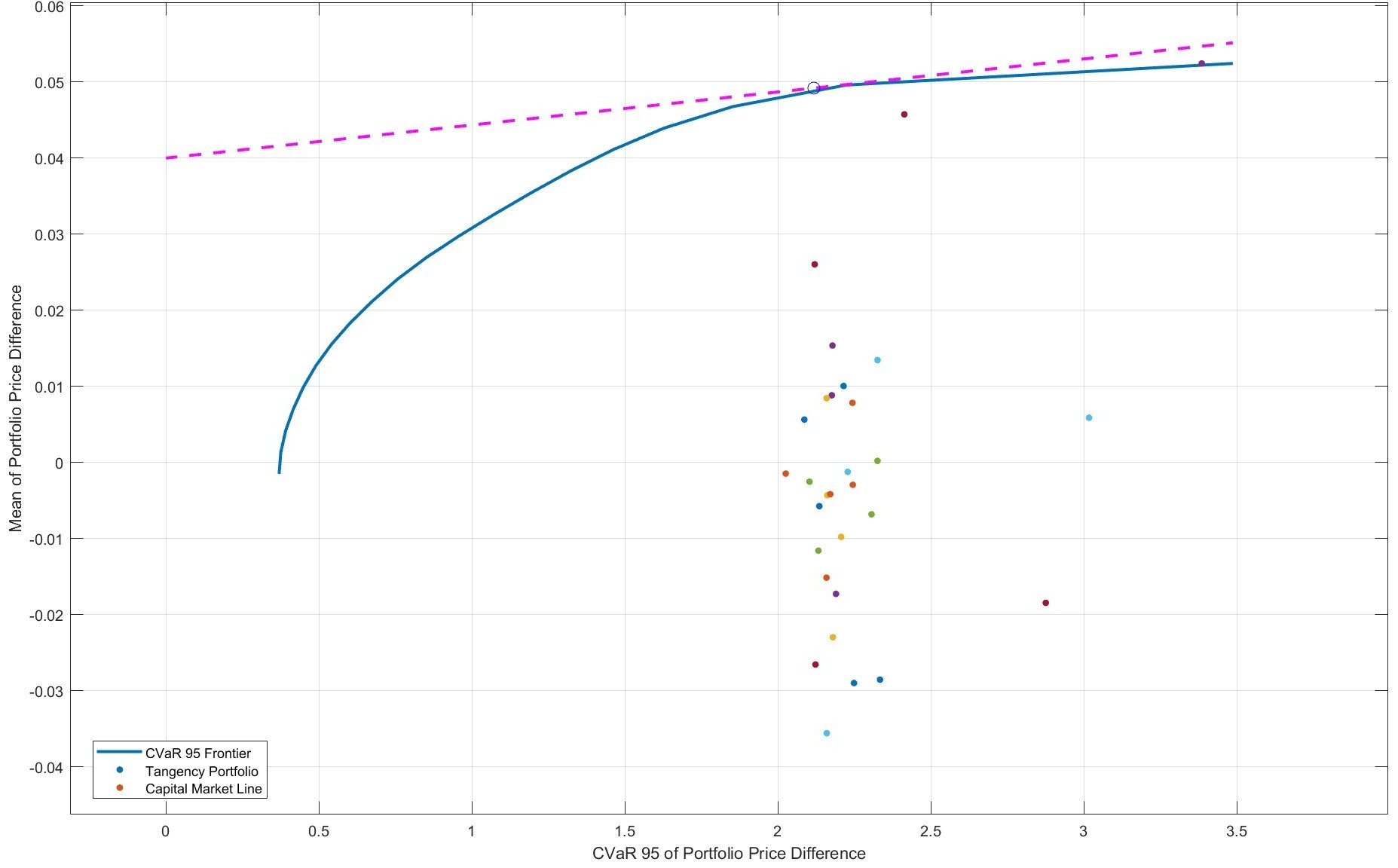} 
        \label{fig:cvar95}
    \end{minipage}
    \hfill
    \begin{minipage}[t]{0.49\textwidth} 
        \centering
        \includegraphics[width=\textwidth]{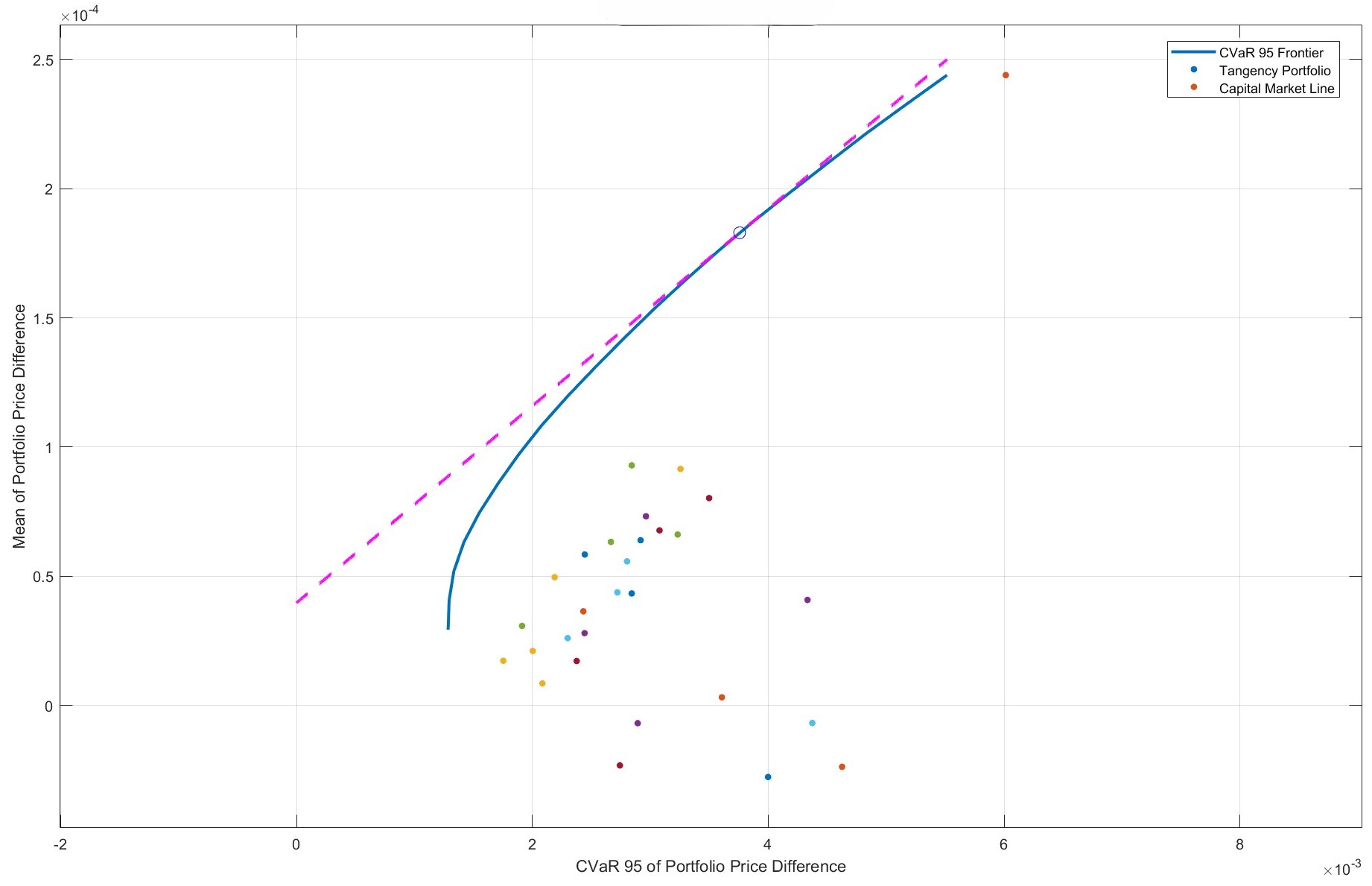} 
        \label{fig:cvar95_mhn}
    \end{minipage}
    \caption{CVaR 95 Optimization-based Efficient Frontier}
    \label{fig:cvar95_EF}
\end{figure}

\begin{figure}[htbp]
    \centering
    \begin{minipage}[t]{0.49\textwidth} 
        \centering
        \includegraphics[width=\textwidth]{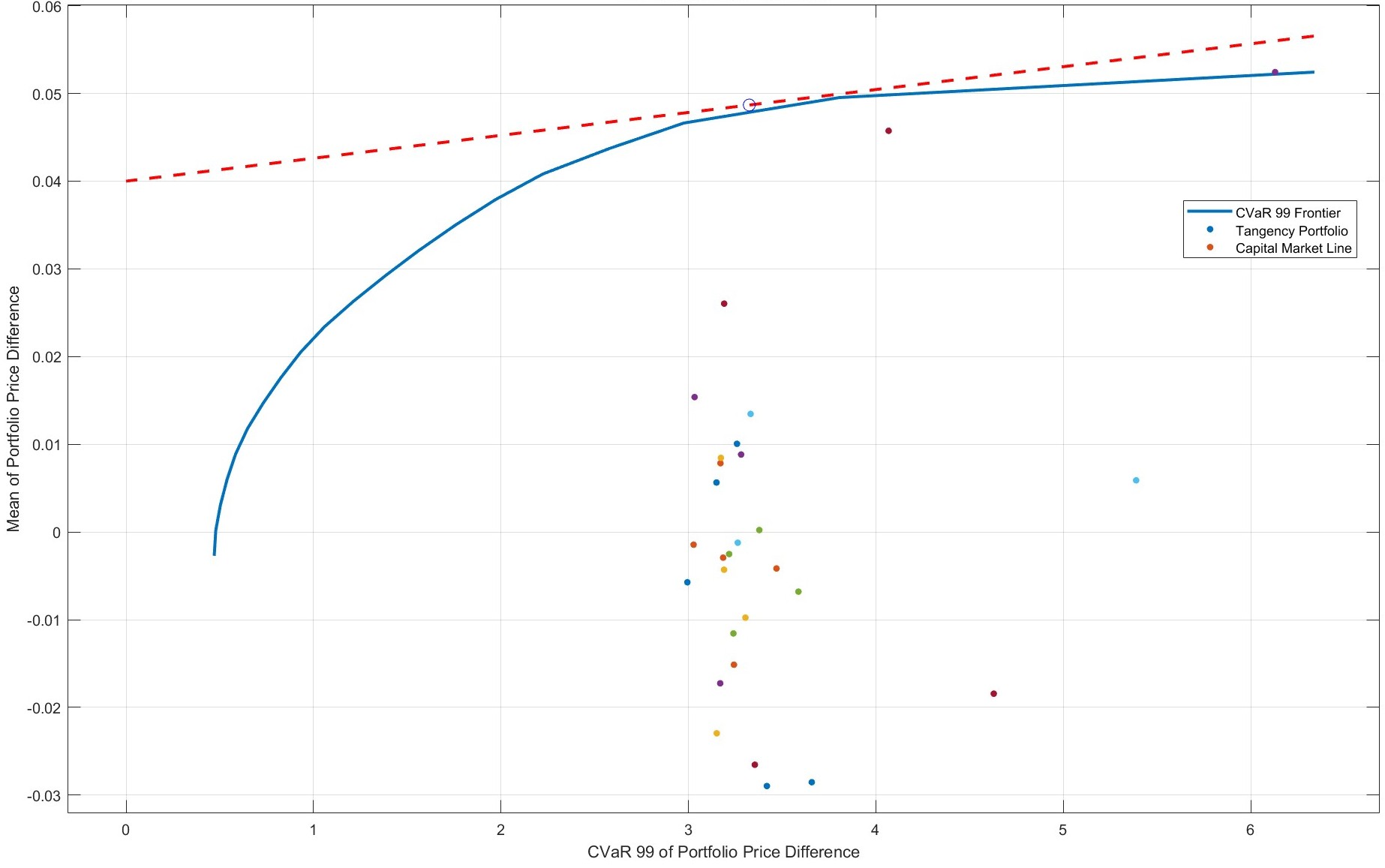} 
        \label{fig:cvar99}
    \end{minipage}
    \hfill
    \begin{minipage}[t]{0.49\textwidth} 
        \centering
        \includegraphics[width=\textwidth]{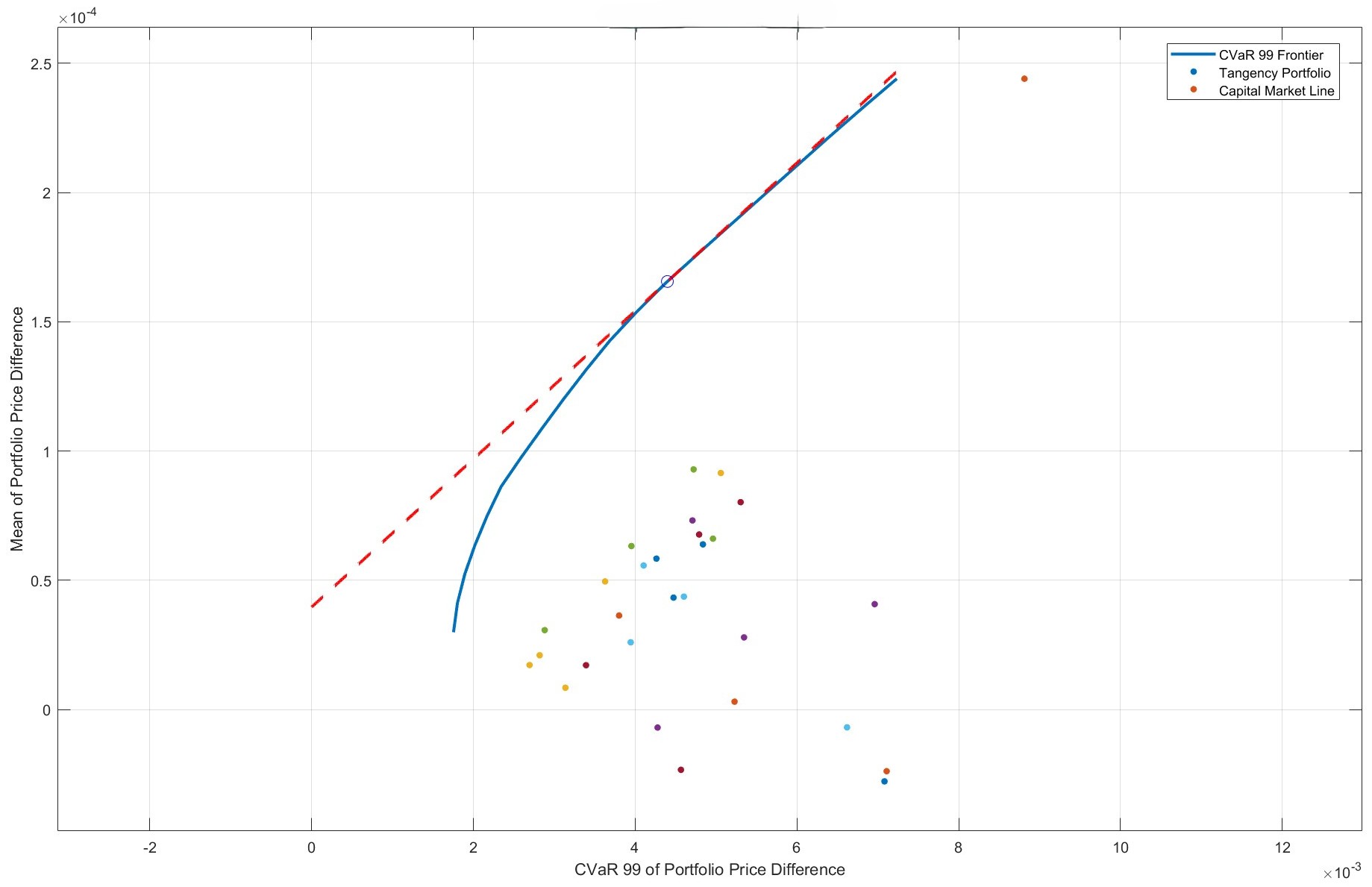} 
        \label{fig:cvar99_mhn}
    \end{minipage}
    \caption{CVaR 99 Optimization-based Efficient Frontier}
    \label{fig:CVaR99_EF}
\end{figure}

Upon solving this problem, the resulting portfolio \(\mathbf{w}_k^{\star}\) achieves the lowest CVaR\(_{\alpha}\) across the given scenarios. Figures~\ref{fig:cvar95_EF} and \ref{fig:CVaR99_EF} plot the efficient frontiers at 95\% and 99\% confidence, respectively, including both ARFIMA-FIGARCH and MHN-GARCH specifications.

\newpage 

\section{Wealth-Equivalent Loss} \label{loss}

Let \(U(W)\) denote the investor’s utility function defined over terminal wealth \(W\).  In this paper we assume a CRRA specification for the representative Dow-30 investor:

\[
U(W) \;=\; 
\begin{cases}
\displaystyle \frac{W^{1-\gamma}}{1-\gamma}, & \gamma \neq 1,\\[1em]
\ln(W), & \gamma = 1,
\end{cases}
\]

\noindent where \(\gamma>0\) is the coefficient of relative risk aversion.  

\noindent Suppose under the benchmark strategy (e.g., the Merton constant‐mix strategy) an investor’s expected utility is

\[
\overline{U_{\mathrm{Merton}}}
\;=\;
\mathbb{E}\bigl[\,U(W_{\mathrm{Merton}})\bigr].
\]

\noindent Under our proposed MHN‐GARCH–based allocation the expected utility is

\[
\overline{{U(W)}^{*}}
\;=\;
\mathbb{E}\bigl[\,{U(W)}^{*}\bigr].
\]

\noindent The \emph{Wealth‐Equivalent Loss} (WEL) is defined as the percentage reduction in initial wealth that, if applied to the proposed strategy, yields the same expected utility as the benchmark:

\[
\overline{{U(W)}^{*}}
\;=\;
\mathbb{E}\!\Bigl[\,U\bigl((1 - \mathrm{WEL})\,W_0 \;\times\; R^*\bigr)\Bigr]
\;=\;
\overline{U_{\mathrm{Merton}}},
\]

\noindent where \(W_0\) is the common initial wealth and \(R^*\) is the gross return factor under the optimal strategy. Solving for \(\mathrm{WEL}\) gives

\[
\mathrm{WEL}
\;=\;
1
\;-\;
\Bigl[
U^{-1}\bigl(\bar U_{\mathrm{Merton}}\bigr)
/W_0
\Bigr]
\]

\noindent where \(U^{-1}(\cdot)\) denotes the inverse of the utility function.  In the CRRA case,

\[
U^{-1}(u)
\;=\;
\begin{cases}
\bigl((1-\gamma)\,u\bigr)^{\frac{1}{1-\gamma}}, & \gamma\neq 1,\\[8pt]
\exp(u), & \gamma=1.
\end{cases}
\]

\noindent A WEL of \(x\)\% means that the proposed strategy would need to start with \((1-x)\)\% of the initial wealth of the benchmark in order to deliver the same expected utility. A positive WEL indicates that the proposed strategy delivers strictly higher expected utility than the benchmark, since one could afford to give up \(x\)\% of initial wealth and remain indifferent. Conversely, a negative WEL implies the proposed strategy is inferior in terms of expected utility.

\section{Sensitivity Analysis} \label{senstivity}

Let \(w_i(\vartheta)\) denote the optimal portfolio weight on asset \(i\), which is a function of a generic scalar model parameter \(\vartheta\).  In this paper we examine three parameters:  

\[
\vartheta \;\in\;\{\gamma,\;\rho,\;a\},
\]  

where \(\gamma\) is coefficient of relative risk aversion, \(\rho\) is the index–asset correlation parameter, and \(a\) is the correlation–driver loading. We define the local sensitivity of weight \(w_i\) to parameter \(\vartheta\) by the partial derivative  

\[
S_i(\vartheta)
\;=\;
\frac{\partial\,w_i(\vartheta)}{\partial \vartheta}\,.
\]

In practice, \(S_i(\vartheta)\) is estimated numerically via central differences:

\[
S_i(\vartheta)
\;\approx\;
\frac{w_i(\vartheta + \Delta\vartheta)\;-\;w_i(\vartheta - \Delta\vartheta)}{2\,\Delta\vartheta},
\]

with \(\Delta\vartheta\) chosen small relative to the plausible parameter range.

To summarize overall model sensitivity, we report the root‐mean‐square sensitivity (RMSS) across the \(N\) risky assets:

\[
\mathrm{RMSS}(\vartheta)
\;=\;
\sqrt{
  \frac{1}{N}\,\sum_{i=1}^N \bigl[S_i(\vartheta)\bigr]^2
}\,.
\]

This metric captures the average magnitude of weight changes per unit parameter shift. A larger \(\bigl|S_i(\vartheta)\bigr|\) indicates that the optimal allocation to asset \(i\) is highly responsive to small changes in \(\vartheta\). The sign of \(S_i(\vartheta)\) reveals the direction of the effect: \(S_i>0\) means \(w_i\) increases with \(\vartheta\), while \(S_i<0\) means it falls. RMSS\((\vartheta)\) provides a single‐number summary of model stability: a small RMSS implies robust allocations across assets, whereas a large RMSS signals potential over‐sensitivity and risk of parameter misspecification.  

In our numerical experiments, we compute \(S_i(\gamma)\), \(S_i(\rho)\), and \(S_i(a)\) at the calibrated parameter values reported in Table~\ref{tab:combined-parameters}, using \(\Delta\gamma=1\), \(\Delta\rho=0.1\), and \(\Delta a=0.5\).  The resulting RMSS statistics justify the observed robustness to correlation perturbations and the pronounced sensitivity to risk aversion and correlation–driver loading highlighted in Section~\ref{sensitivityanalysis}.

\end{document}